%% file: arxiv_final.tex
\documentclass[amsmath,amssymb,aps,pra,superscriptaddress,floats,twocolumn,longbibliography,10pt]{revtex4-1} %onecolumn,preprint,12pt %twocolumn,10pt

\usepackage{float}
\usepackage{mathrsfs}
\usepackage{bm}
\usepackage{dsfont}
\fontfamily{ptm}\selectfont
\usepackage{stmaryrd}
\usepackage[usenames,dvipsnames]{color}
\usepackage{graphicx,subfigure}
\usepackage{wrapfig}
\usepackage{xcolor}
\usepackage{cancel}
\usepackage{amsmath,amssymb}
\usepackage{times}
\usepackage[latin1]{inputenc}
\usepackage{setspace} %%%%%
\usepackage[normalem]{ulem}
\usepackage[plainpages=false,pdfpagelabels,colorlinks=true,linkcolor=blue,urlcolor=blue,citecolor=blue,pdftitle={Title},pdfauthor={},pdfdisplaydoctitle=true,pdfduplex=DuplexFlipLongEdge]{hyperref}
\usepackage{textcomp}
\usepackage{multirow}
\usepackage{bm}
\usepackage{mathrsfs}

\newcommand{\ket}[1]{\ensuremath{{\vert #1 \rangle}}}
\newcommand{\bra}[1]{\ensuremath{\langle #1 \vert}}
\renewcommand{\vec}[1]{\ensuremath{\mathbf{#1}}}

\newcommand{\abs}[1]{\ensuremath{\left\vert #1 \right\vert}}
\newcommand{\q}{\ensuremath{\vec{q}}}

\renewcommand{\k}{\ensuremath{\vec{k}}}
\newcommand{\KK}{\ensuremath{\vec{K}}}
\newcommand{\KKp}{\ensuremath{\vec{K'}}}
\newcommand{\dd}{\ensuremath{\vec{d}}}

\newcommand{\spinor}[2]{\binom{#1}{#2}}
\newcommand{\braket}[2]{\ensuremath{{\langle #1\vert #2 \rangle}}}
\newcommand{\wel}[3]{\ensuremath{W^{#1}_{\mathbf{#2} \rightarrow \mathbf{#3}}}}
\newcommand{\rop}{\hat{\bold{r}}}

\newcommand{\uket}[2]{\ensuremath{\ket{u^{#1}_{#2}}}}
\newcommand{\ubra}[2]{\ensuremath{\bra{u^{#1}_{#2}}}}

\newcommand{\nbands}{\ensuremath{\mathcal{N}}}

\newcommand{\nocontentsline}[3]{}
\newcommand{\tocless}[2]{\bgroup\let\addcontentsline=\nocontentsline#1{#2}\egroup}

\usepackage{ulem}

\newcommand{\editout}[1]{{\bf \sout{#1}}}
\renewcommand{\editout}[1]{}

\begin{document}

\singlespacing
\noindent
\title{Bloch state tomography using Wilson lines}

\newcommand{\lmu}{Fakult\"{a}t f\"{u}r Physik, Ludwig-Maximilians-Universit\"{a}t M\"{u}nchen, Schellingstr.\ 4, 80799 Munich, Germany}
\newcommand{\mpq}{Max-Planck-Institut f\"{u}r Quantenoptik, Hans-Kopfermann-Str.\ 1, 85748 Garching, Germany}
\newcommand{\stanford}{Department of Physics, Stanford University, Stanford, California 94305, USA}
\newcommand{\harvard}{Department of Physics, Harvard University, Cambridge, Massachusetts 02138, USA}
\newcommand{\cambridge}{Cavendish Laboratory, University of Cambridge, J. J. Thomson Avenue, Cambridge CB3 0HE, UK}
\newcommand{\kais}{Department of Physics and Research Center OPTIMAS, University of Kaiserslautern, Germany}
\newcommand{\kaiserslautern}{Graduate School Materials Science in Mainz, Gottlieb-Daimler-Strasse 47, 67663 Kaiserslautern, Germany}
\newcommand{\caltech}{Institute for Quantum Information and Matter, Department of Physics, California Institute of Technology, Pasadena, CA 91125, USA}
\newcommand{\mail}{E-mail: }

\noindent
\author{T.~\ Li}
\affiliation{\lmu}
\affiliation{\mpq}

\author{L.~\ Duca}
\affiliation{\lmu}
\affiliation{\mpq}

\author{M.~\ Reitter}
\affiliation{\lmu}
\affiliation{\mpq}

\author{F.~\ Grusdt}
\affiliation{\kais}
\affiliation{\kaiserslautern}
\affiliation{\harvard}

\author{E.~\ Demler}
\affiliation{\harvard}

\author{M.~\ Endres}
\affiliation{\harvard}
\affiliation{\caltech}

\author{M.~\ Schleier-Smith}
\affiliation{\stanford}

\author{I.~\ Bloch}
\affiliation{\lmu}
\affiliation{\mpq}

\author{U.\ Schneider}
\affiliation{\lmu}
\affiliation{\mpq}
\affiliation{\cambridge}

%\affiliation{\mail}
\begin{abstract}
Topology and geometry are essential to our understanding of modern physics, underlying many foundational concepts from high energy theories, quantum information, and condensed matter physics. In condensed matter systems, a wide range of phenomena stem from the geometry of the band eigenstates, which is encoded in the matrix-valued Wilson line for general multi-band systems. Using an ultracold gas of Rb atoms loaded in a honeycomb optical lattice, we realize strong-force dynamics in Bloch bands that are described by Wilson lines and observe an evolution in the band populations that directly reveals the band geometry. Our technique enables a full determination of band eigenstates,  Berry curvature, and topological invariants, including single- and multi-band Chern and $\mathds{Z}_2$ numbers.
\end{abstract}
\maketitle

Geometric concepts play an increasingly important role in elucidating the behavior of condensed matter systems. In band structures without degeneracies, the geometric phase acquired by a quantum state during adiabatic evolution elegantly describes a spectrum of phenomena~\cite{Xiao2010}. This geometric phase\textemdash known as the Berry phase\textemdash is used to formulate the Chern number~\cite{Thouless1982}, which is the topological invariant characterizing the integer quantum Hall effect~\cite{Klitzing1980}. However, condensed matter properties that are determined by multiple bands with degeneracies, such as in topological insulators~\cite{Hasan2010,Qi2011}  and graphene~\cite{Neto2009}, can seldom be understood with standard Berry phases. Recent work has shown that such systems can instead be described using Wilson lines~\cite{Yu2011,Alexandradinata2014b,Alexandradinata2014,Grusdt2014}.

\begin{figure}[t!]
	\centering
		\includegraphics[width=87mm]{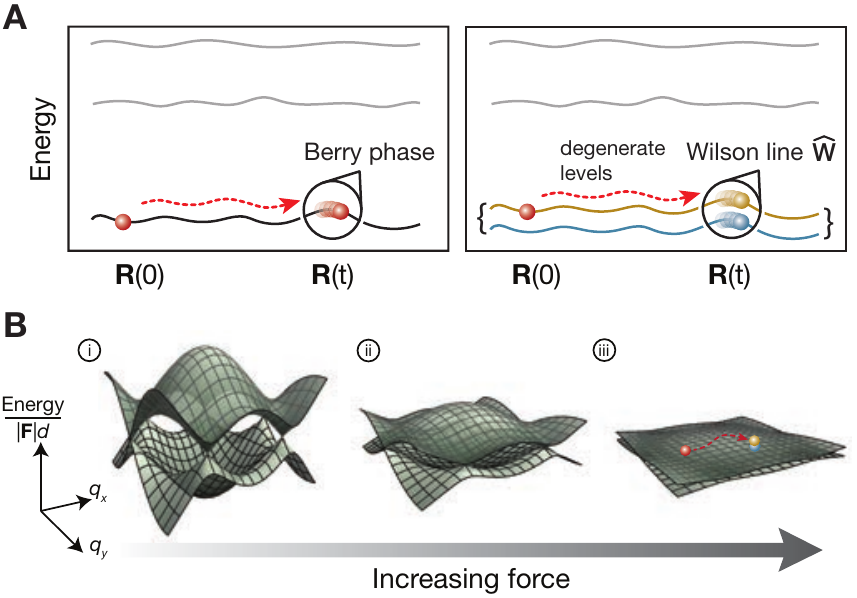}
    \caption{\textbf{Wilson lines and effectively degenerate Bloch bands. (A)} In a non-degenerate system (left), adiabatic evolution of a state through parameter space $\vec{R}$ results in the acquisition of a geometric phase factor, known as the Berry phase. In a degenerate system (right), the evolution is instead governed by a matrix-valued quantity called the Wilson line. If the degenerate levels can be experimentally distinguished (blue and yellow colouring), then population changes between the levels are detectable.
    \textbf{(B)} The band structure of the lowest two bands of the honeycomb lattice in effective energy units of $\smash{\abs{\vec{F}}d}$, where $\vec{F}$ is the applied force used to transport the atoms and $d$ is the distance between nearest-neighbor lattice sites. As $\abs{\mathbf{F}}$ is increased, the largest energy scale of the bands becomes small compared to $\smash{\abs{\mathbf{F}}d}$. At large forces (iii), the effect of the band energies is negligible and the system is effectively degenerate. In this regime, the evolution is governed by the Wilson line operator. We distinguish between the bands using a band mapping technique that detects changes in the band population along the Wilson line path.}
	\label{Fig:1}
\end{figure}

Wilson lines encode the geometry of degenerate states~\cite{Wilczek1984}, providing indispensable information for the ongoing effort to identify the topological structure of bands. For example, the eigenvalues of Wilson-Zak loops (i.e., Wilson lines closed by a reciprocal lattice vector) can be used to formulate the $\mathds{Z}_2$ invariant of topological insulators~\cite{Yu2011} and identify topological orders protected by lattice symmetries~\cite{Alexandradinata2014,Alexandradinata2014b}. Although experiments have accessed the geometry of isolated bands through various methods including transport measurements~\cite{Klitzing1980,Jotzu2014,Aidelsburger2014}, interferometry~\cite{Duca2015,Atala2013}, and angle-resolved photoemission spectroscopy \cite{Hasan2015,Qi2011}, Wilson lines have thus far remained a theoretical construct~\cite{Yu2011,Alexandradinata2014b,Alexandradinata2014,Grusdt2014}.

Using ultracold atoms in a graphene-like honeycomb lattice, we demonstrate that Wilson lines can be accessed and used as versatile probes of band structure geometry. Whereas the Berry phase merely multiplies a state by a phase factor, the Wilson line is a matrix-valued operator that can mix state populations~\cite{Wilczek1984}~(Fig.\ 1A). We measure the Wilson line by detecting changes in the band populations~\cite{Greiner2001} under the influence of an external force, which transports atoms through reciprocal space~\cite{Dahan1996}. In the presence of a force $\mathbf{F}$, atoms with initial quasimomentum $\mathbf{q}(0)$ evolve to quasimomentum $\mathbf{q}(t)=\mathbf{q}(0)+\mathbf{F}t/\hbar$ after a time $t$. If the force is sufficiently weak and the bands are non-degenerate, the system will undergo adiabatic Bloch oscillations and remain in the lowest band~\cite{Dahan1996}. In this case, the quantum state merely acquires a phase factor comprised of the geometric Berry phase and a dynamical phase. At stronger forces, however, transitions to other bands occur and the state evolves into a superposition over several bands.

When the force is infinite with respect to a chosen set of bands, the effect of the dispersion vanishes and the bands appear as effectively degenerate~(Fig.\ 1B). The system then evolves according to the formalism of Wilczek and Zee for adiabatic motion in a degenerate system~\cite{Wilczek1984}. 
The unitary time-evolution operator describing the dynamics is the Wilson line matrix~\cite{SOM}:
\begin{align}\label{Wilson_def}
\hat{\mathbf{W}}_{\mathbf{q}(0) \rightarrow \mathbf{q}(t)}=\mathcal{P}\text{exp}[i\int_\mathcal{C}d\mathbf{q} \hat{\mathbf{A}}_\mathbf{q}],
\end{align}
where the path-ordered ($\mathcal{P}$) integral runs over the path $\mathcal{C}$ in reciprocal space from $\vec{q}(0)$ to $\mathbf{q}(t)$ and $\mathbf{\hat{A}_q}$ is the Wilczek-Zee connection, which encodes the local geometric properties of the state space. 

In a lattice system with Bloch states $\ket{\Phi_{\mathbf{q}}^{n}} = e^{i\vec{q}\cdot\mathbf{\hat{r}}} \ket{u_{\mathbf{q}}^{n}}$ in the $n^\text{th}$ band at quasimomentum $\mathbf{q}$, where $\mathbf{\hat{r}}$ is the position operator, the elements of the Wilczek-Zee connection are determined by the cell-periodic part $\ket{u_{\mathbf{q}}^{n}}$ as $\mathbf{A}^{n,n'}_{\mathbf{q}}~=~i \bra{u_{\mathbf{q}}^n}\nabla_\q \ket{u_{\mathbf{q}}^{n'}}$. The diagonal elements ($n=n'$) are the Berry connections of the individual Bloch bands, which yield the Berry phase when integrated along a closed path. The off-diagonal elements ($n \neq n'$) are the inter-band Berry connections, which couple the bands and induce inter-band transitions.

\begin{figure}[!t]
	\centering
		\includegraphics[width=87mm]{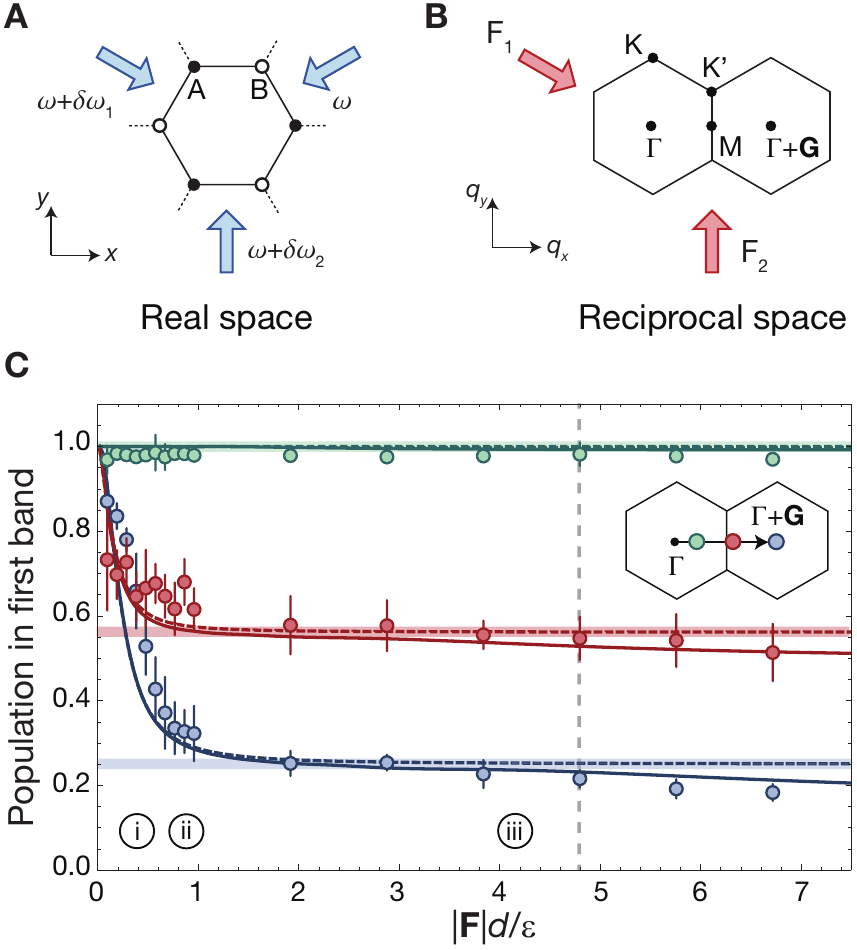}
    \caption{\textbf{Reaching the Wilson line regime in the honeycomb lattice. (A)} Schematic of the honeycomb lattice in real space with  A (B) sublattice sites denoted by solid (open) circles. A static lattice is formed by interfering three in-plane laser beams (blue arrows) with frequency $\omega$. Sweeping the frequency of beam $i$ by $\delta \omega_{i}$  creates a force $\mathbf{F_{i}}$ in the lattice frame in the propagation direction of beam $i$~\cite{SOM}. \textbf{(B)} Two copies of the first BZ of the honeycomb lattice, separated by a reciprocal lattice vector~$\mathbf{G}$. By changing the relative strengths of $\mathbf{F_{i}}$ (red arrows), the atoms can be moved along arbitrary paths in reciprocal space. Each BZ features non-equivalent Dirac points $\mathbf{K}$ and $\mathbf{K'}$ at the corners of the hexagonal cell. High-symmetry points $\Gamma$, at the center of the BZ, and M, at the edge of the BZ, are also shown. \textbf{(C)} The population remaining in the first band for different forces after transport to $\Gamma+0.2\mathbf{G}$ (green), $\Gamma+0.55\mathbf{G}$ (red), and $\Gamma+\mathbf{G}$ (blue). Inset numbers i to iii refer to band schematics in Fig.\ 1B, representing the diminishing effect of the dispersion for increasing force. The data agree well with a two-level, tight-binding model (dashed line) which approaches the Wilson line regime (thick shaded line) at large forces. Discrepancies at larger forces result from transfer to higher bands and match well with \textit{ab initio} theory using a full band structure calculation including the first six bands (thin solid line)~\cite{SOM}. For all subsequent data, we use $\abs{\mathbf{F}}d/\varepsilon=4.8$, indicated by the dashed gray line. Error bars indicate the standard error of the mean from ten shots per data point. 
}
	\label{Fig:2}
\end{figure}

Although the evolution described by Eq.~\ref{Wilson_def} must be path-ordered when the Wilczek-Zee connections at different quasimomenta do not commute, it can also be path-independent under certain circumstances~\cite{Zwanziger1990,Alexandradinata2014}. For example, when the relevant bands span the same Hilbert space for all quasimomenta, as is the case in our system, the Wilson line operator describing transport of a Bloch state from $\vec{Q}$ to $\vec{q}$ reduces to $\vec{\hat{W}}_{\vec{Q}\rightarrow\vec{q}}=e^{i(\vec{q}-\vec{Q})\cdot\hat{\vec{r}}}$~ ~\cite{SOM,Alexandradinata2014,Kingsmith1993}. Consequently, the elements of the Wilson line operator simply measure the overlap between the cell-periodic Bloch functions at the initial and final quasimomenta~\cite{Alexandradinata2014,Kingsmith1993}:
\begin{align}\label{eq:overlap}
W^{mn}_{\mathbf{Q} \rightarrow \mathbf{q}}=\bra{\Phi_\mathbf{{q}}^m}e^{i (\vec{q}-\vec{Q}) \cdot \hat{\vec{r}}}\ket{\Phi_{\vec{Q}}^n} = \braket{u_\mathbf{q}^m}{u_\mathbf{Q}^n}.
\end{align}
Hence, access to the Wilson line elements facilitates the characterization of band structure topology in both path-dependent and path-independent evolution. In both cases, the topological information is encoded in the eigenvalues of the Wilson-Zak loops. In the latter case, the simplified form of the Wilson line in Eq.~\ref{eq:overlap} additionally enables a map of the cell-periodic Bloch functions over the entire BZ in the basis of the states $\ket{u_\mathbf{Q}^n}$ at the initial quasimomentum $\vec{Q}$.

We create the honeycomb lattice by interfering three blue-detuned laser beams at 120(1)$^\circ$ angles~(Fig.\ 2A).  At a lattice depth $V_0$=5.2(1)$E_r$, where $E_r$=h$^2/(2m\lambda_L^2)$ is the recoil energy, $\lambda_L$ is the laser wavelength, and $m$ is the mass of $^{87}$Rb, the combined width $\varepsilon\approx \mathrm{h}\times3$~kHz of the lowest two bands is much smaller than the $\mathrm{h}\times15$~kHz gap to higher bands. Consequently, there exists a regime of forces where transitions to higher bands are suppressed and the system is well-approximated by a two-band model~\cite{SOM}. 

We probe the lattice geometry with a nearly pure Bose-Einstein condensate of $^{87}$Rb, which is initially loaded into the lowest band at quasimomentum $\mathbf{q}=\Gamma$, the center of the Brillouin zone~(BZ) (Fig.\ 2B).  To move the atoms in reciprocal space, we linearly sweep the frequency of the beams to uniformly accelerate the lattice, thereby generating a constant inertial force in the lattice frame. By independently controlling the frequency sweep rate of two beams (see~Fig.\ 2A), we can tune the magnitude and direction of the force and move the atoms along arbitrary paths in reciprocal space.

\begin{figure}[!t]
	\centering
		\includegraphics[width=84mm]{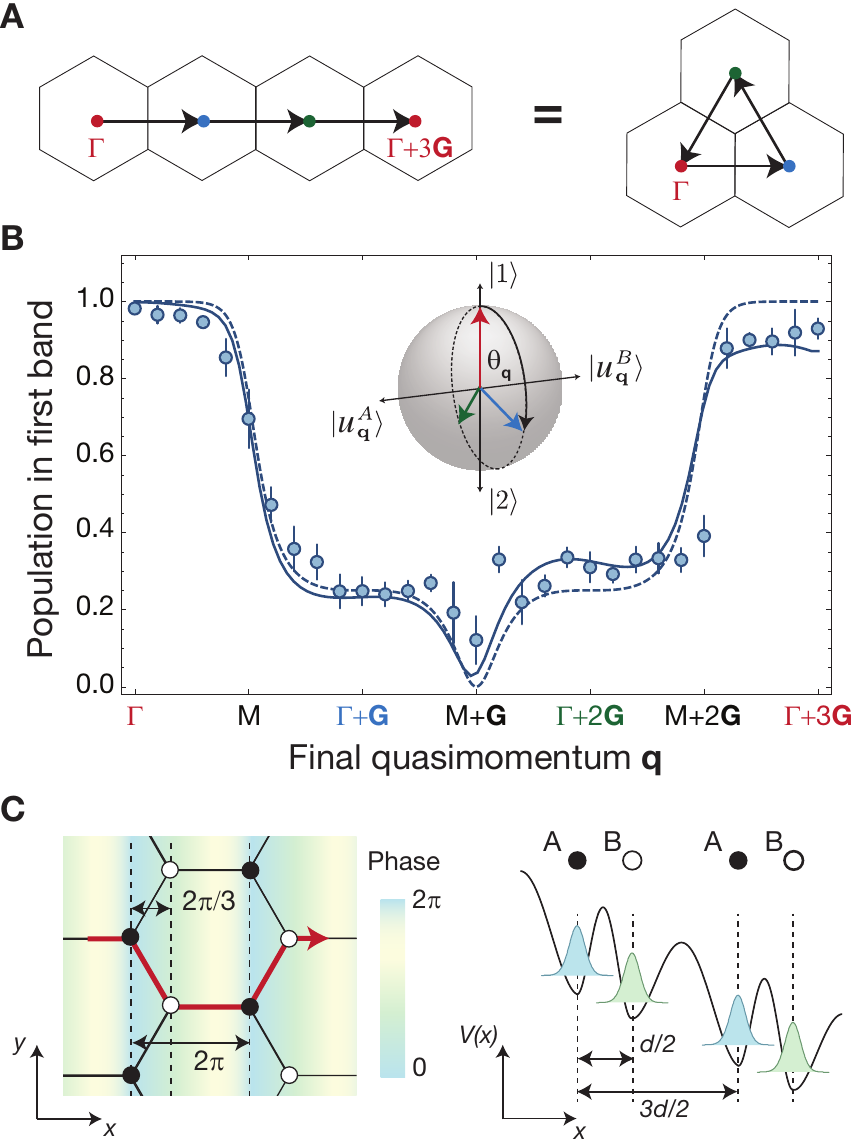}
\caption{\textbf{Measuring mixing angles $\theta_\mathbf{q}$ at different quasimomenta. (A)} Because of the three-fold-rotational symmetry of the honeycomb lattice, a path from $\Gamma$ to $\Gamma+3\mathbf{G}$  is equivalent to a triangle-shaped path with each leg of length $\abs{\mathbf{G}}$, beginning and ending at $\Gamma$. Coloured dots correspond to coloured quasimomentum labels in (B).
\textbf{(B)} The population remaining in the first band after transport to final quasimomentum $\mathbf{q}$. Theory lines are a single-particle solution to the dynamics using a full lattice potential and including the first six bands (solid) and a two-band, tight-binding model (dashed)~\cite{SOM}. The inset Bloch sphere depicts the transported state at $\Gamma$ (red), $\Gamma+\mathbf{G}$ (blue), and $\Gamma+2\mathbf{G}$ (green) in the basis of the cell-periodic Bloch functions at $\Gamma$.  Error bars represent the standard error of the mean from averaging 9-11 shots, with the exception of $\mathbf{q}=\text{M}+\mathbf{G}$ and $\mathbf{q}=\text{M}+0.9\mathbf{G}$, which show the average of 20 shots. 
\textbf{(C)} Transport of a Bloch state by one reciprocal lattice vector corresponds to a $2\pi$ phase shift in the real-space wavefunctions of each sublattice site. Projecting the combined lattice and gradient potential \textit{V(x)} along the path shown in red onto the \textit{x}-axis, which is the direction of the applied force in the measurements of Fig.\ 2B and Fig.\ 3B, highlights the effect of the real-space embedding of the honeycomb lattice. Since the distance between A (solid circles) and B sites (open circles) is 1/3 the distance between sites of the same type, there is a phase difference of $2\pi/3$ between the real-space wavefunctions of A and B sites, which gives rise to the band mixing.
}
	\label{Fig:3}
\end{figure}

To verify that we can access the Wilson line regime, where the dynamics are governed entirely by geometric effects, we transport the atoms from $\Gamma$ to different final quasimomenta using a variable force $\abs{\mathbf{F}}$ and perform band mapping~\cite{Greiner2001} to measure the population remaining in the lowest band (Fig.\ 2C). For vanishing forces, we recover the adiabatic limit, where the population remains in the lowest band. For increasing forces (i and ii in Fig.\ 1B), where the gradient $\abs{\mathbf{F}}d$ over the distance between A and B sites $d$ is less than the combined width $\varepsilon$, the population continuously decreases. However, at strong forces (iii in Fig.\ 1B), where $\smash{\abs{\mathbf{F}}d>\varepsilon}$, the population saturates at a finite value. For example, after transport by one reciprocal lattice vector (blue data in Fig.\ 2C), about one quarter of the atoms remain in the first band, in stark contrast to typical Landau-Zener dynamics, where the population vanishes for strong forces~\cite{Shevchenko2010}.

Theoretically, the population in the first band after the strong-force transport directly measures the Wilson line element $\abs{W^{11}_{\Gamma \rightarrow \mathbf{q}}}^2=\abs{\bra{\Phi_\mathbf{{q}}^1}\hat{\mathbf{W}}_{\Gamma \rightarrow \mathbf{q}}\ket{\Phi_{\Gamma}^1}}^2$ in the basis of the band eigenstates. Based on Eq.~\ref{eq:overlap}, the saturation value $\abs{W^{11}_{\Gamma \rightarrow \mathbf{q}}}^2=\abs{\braket{u_\vec{q}^1}{u_\Gamma^1}}^2$ of the population after transport to $\mathbf{q}$ is a measure of the overlap between the cell-periodic Bloch functions of the first band $\ket{u_\q^1}$ at $\Gamma$ and $\mathbf{q}$. Notably, for the case of transport by one reciprocal lattice vector $\mathbf{G}$, the cell-periodic parts $\ket{u_\mathbf{q}^n}$ are not identical, despite the unity overlap of the Bloch states $\ket{\Phi_{\mathbf{q}}^{n}}$ at $\Gamma$ and $\Gamma+\mathbf{G}$.  In contrast to the typical Landau-Zener case, they are also not orthogonal---hence the finite saturation value.

To corroborate that our experiment measures the Wilson line, we transport atoms initially in the ground state at $\Gamma$ by up to three reciprocal lattice vectors (Fig.\ 3). The three-fold rotational symmetry of the lattice, combined with the symmetry of its \textit{s-}orbitals, makes the path from $\Gamma$ to $\Gamma+3\vec{G}$  equivalent to the triangular path shown in Fig.\ 3A, such that the overlap between cell-periodic components of the Bloch wavefunctions at the two endpoints is unity (see Eq.~\ref{eq:overlap}). Correspondingly, we expect to recover all population in the lowest band after transport from $\Gamma$ to $\Gamma+3\vec{G}$.  This prediction is confirmed in Fig.\ 3B, where we plot the population remaining in the first band after transport to final quasimomentum $\vec{q}$.  The data are well described by a tight-binding model that takes into account the relative phase between orbitals on A and B sites of the lattice due to the Wilson line $\hat{\mathbf{W}}_{\vec{\Gamma} \rightarrow \vec{q}} = e^{i \vec{q}\cdot \hat{\vec{r}}}$. Physically, this can be  understood by assuming that the real-space wavefunction simply accumulates a position-dependent phase when the strong force $\vec{F}~=~\hbar(\vec{q}-\vec{Q})/t$ is applied for a short time~$t$~(Fig.\ 3C). Notably, the result depends crucially on the real-space embedding of the lattice and would be different in, e.g., a brick-wall incarnation~\cite{Tarruell2012} of the same tight-binding model. Discrepancies from the tight-binding model result from population transfer to higher bands~\cite{SOM}.

As the Wilson line enables a comparison of the cell-periodic Bloch functions at any two quasimomenta (Eq.~\ref{eq:overlap}), it can in principle be applied to fully reconstruct these states throughout reciprocal space. To this end, it is convenient to represent the state $\ket{u_\mathbf{q}^1}$ at quasimomentum $\mathbf{q}$ in the basis of cell-periodic Bloch functions $\ket{1}=\ket{u_\mathbf{Q}^1}$ and $\ket{2}=\ket{u_\mathbf{Q}^2}$ at a fixed reference quasimomentum $\mathbf{Q}$ as 
\begin{align}\label{eq:thetaphi} \ket{u_\mathbf{q}^1}=\cos\frac{\theta_\mathbf{q}}{2}\ket{1}+\sin\frac{\theta_\mathbf{q}}{2}e^{i\phi_\mathbf{q}}\ket{2}.
\end{align}
Mapping out the geometric structure of the lowest band therefore amounts to obtaining $\theta_\mathbf{q}$ and $\phi_\mathbf{q}$, which parametrize the amplitude and phase of the superposition between the reference Bloch states, for each quasimomentum $\vec{q}$~\cite{Hauke2014,Alba2011}. Whereas the total phase of $\ket{u_\mathbf{q}^1}$ is gauge dependent, i.e., it can be chosen for each $\mathbf{q}$, the relative phase $\phi_\mathbf{q}$ is fixed for all $\mathbf{q}$ once the basis states $\ket{1}$ and $\ket{2}$ are fixed. Throughout this work, we choose the basis states at reference point $\vec{Q}=\Gamma$.

In this framework, the population measurements in Fig.\ 3B constitute a reconstruction of the mixing angle $\theta_\vec{q}=2\arccos\abs{W^{11}_{\Gamma \rightarrow \mathbf{q}}}$. This can be visualized as the rotation of a pseudospin on a Bloch sphere, where the north (south) pole represents $\ket{1}$ $(\ket{2})$. As a function of quasimomentum $\vec{q}$, the angle $\theta_\vec{q}$ winds by $2\pi/3$ per reciprocal lattice vector  (see inset of Fig.\ 3B).  

To obtain the relative phase $\phi_\vec{q}$, which is directly connected to the Wilson line via $\phi_\mathbf{q}=\text{Arg}[\wel{11}{\mathbf{Q}}{\mathbf{q}}]-\text{Arg}[\wel{12}{\mathbf{Q}}{\mathbf{q}}]$~\cite{SOM}, we perform a procedure analogous to Ramsey or St{\"u}ckelberg interferometry~\cite{Zenesini2010,Kling2010}. We initialize atoms in the lowest band at $\Gamma-\mathbf{G}$ and rapidly transport them by one reciprocal lattice vector to prepare a superposition of band eigenstates at the reference point $\Gamma$ (i in Fig.\ 4A).  We then hold the atoms at $\Gamma$ for a variable time (ii), during which the phase of the superposition state precesses at a frequency set by the energy difference between the bands at $\Gamma$. Following this preparation sequence, we rapidly transport the superposition state to a final quasimomentum $\vec{q}_\alpha$, lying at angular coordinate $\alpha$ on a circle of radius $\abs{\vec{G}}$ centered at $\Gamma$. Measuring the population of the first band as a function of hold time yields an interference fringe that reveals the relative phase $\phi_\alpha$~\cite{SOM}.

 \begin{figure}[htb]
	\centering
		\includegraphics[width=84mm]{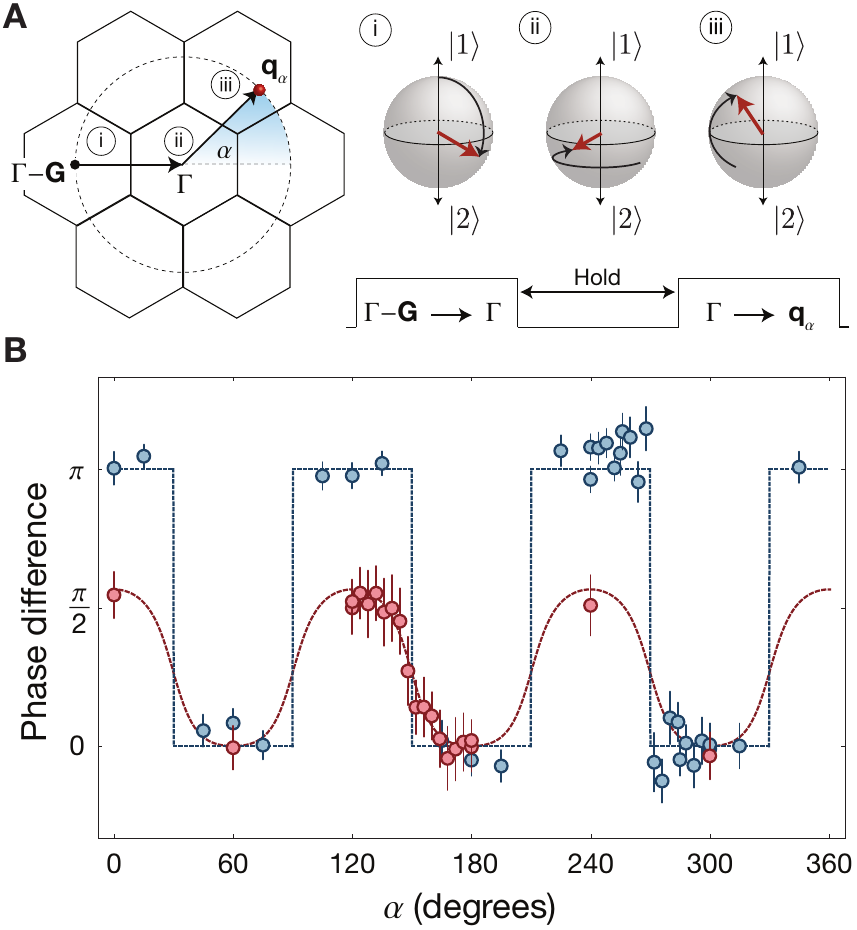}
    \caption{\textbf{Measuring relative phases $\phi_\mathbf{q}$ at different quasimomenta. (A)} Schematic of the interferometric sequence in the extended BZ scheme (left) and the corresponding rotation on the Bloch sphere (right). To create a superposition state, atoms initially in the lowest eigenstate at $\Gamma-\mathbf{G}$ are rapidly transported to $\Gamma$ (i). The phase of the superposition state is controlled by varying the hold time at $\Gamma$ (ii). After the state preparation, the atoms are transported to a final quasimomentum $\mathbf{q_\alpha}$, which is parametrized by the angle $\alpha$ and lies on a circle of radius $\abs{\vec{G}}$ centered at $\Gamma$  (iii). \textbf{(B)} Phases $\phi_\alpha$ referenced to $\alpha$=180$^\circ$ for the lattice with AB-site degeneracy (blue) and AB-site offset (red). Data in blue have been offset by +120$^\circ$ for visual clarity. Dashed lines are a two-band, tight-binding calculation with $\Delta/J$=0 (blue) and $\Delta/J=3.1$ (red), where $\smash{J=\text{h}\times500(10)}~$Hz. Error bars indicate fit errors. 
}
	\label{Fig:4}
\end{figure}

We observe quantized jumps of $\pi$ in the phase of the interference fringe each time $\alpha$ is swept through a Dirac point, i.e., every $60^\circ$ (blue circles in Fig.\ 4B)~\cite{Lim2014,Lim2015}. The binary nature of the phases is a consequence of the degeneracy between A and B sites, which dictates that the band eigenstates at each quasimomentum be an equal superposition of states $\ket{\Phi_\vec{q}^A}$ and $\ket{\Phi_\vec{q}^B}$ on the A and B sublattices~\cite{SOM}. Therefore, on the Bloch sphere, the pseudospin is constrained to rotate on a meridian about an axis whose poles represent the corresponding cell-periodic functions $\ket{u_\vec{q}^A}$ and $\ket{u_\vec{q}^B}$ (inset of Fig.\ 3B). When we remove this constraint by introducing an energy offset between A and B sites~\cite{SOM,Baur2014}, we observe smoothly varying phases that are always less than $\pi$ (red circles in Fig.\ 4B). The dependence of the phase on angle $\alpha$ indicates both the breaking of inversion symmetry and the preservation of the three-fold rotational symmetry of the lattice.

Apart from reconstructing the cell-periodic Bloch functions, our method also provides access to eigenvalues of Wilson-Zak loops, $\hat{\mathbf{W}}_{\q\rightarrow\q+\mathbf{G}}$, which is essential for determining various topological invariants~\cite{Yu2011,Alexandradinata2014, Alexandradinata2014b}. To this end, we split the Wilson-Zak-loop matrix into a global phase factor, which can be measured by extending previous methods~\cite{Aidelsburger2014,Atala2013, Duca2015, Abanin2013}, and an $SU(2)$ matrix with eigenvalues $e^{\pm i \xi}$. Using the data from  Fig.\ 3B and Fig.\ 4B, we reconstruct the eigenvalues for a loop transporting from $\Gamma$ to $\Gamma+\mathbf{G}$, up to multiples of $\pi$~\cite{SOM}. We find the eigenvalue phases to be $\xi=1.03(2)\pi/3$, in good agreement with the value of $\pi/3$ predicted from the two-band model. Remarkably, we measure the same eigenvalues even when the band eigenstates are modified by an energy offset between A and B sites~\cite{SOM}. This invariance is a direct consequence of the real-space representation of the Wilson-Zak loop, $\hat{\mathbf{W}}_{\Gamma\rightarrow\Gamma+\mathbf{G}}=e^{i \vec{G} \cdot \hat{\vec{r}}}$ (see Eq.~\ref{eq:overlap} and~\cite{SOM}). Because the Wilson-Zak loop depends only on the position operator~$\rop$, the eigenvalues are determined solely by the physical locations of the lattice sites, which are unchanged by the energy offset.

Our versatile approach to measuring Wilson lines only employs standard techniques that are broadly applicable in ultracold atom experiments, and can be extended to higher numbers of bands by adopting ideas from quantum process tomography~\cite{Poyatos1997}. Provided that the relevant local Hilbert space is identical for all quasimomenta, our method provides a complete map of the eigenstates over the BZ, giving access to the Berry curvature and Chern number. When this is not the case, the same techniques enable the reconstruction of Wilson-Zak loop eigenvalues, which directly probe the geometry of the Wannier functions~\cite{Alexandradinata2014} and, therefore, the polarization of the system~\cite{Kingsmith1993,Soluyanov2011}. Consequently, these eigenvalues can reveal the topology of bands with path-dependent and non-Abelian Wilson lines~\cite{Wilczek1984,Alexandradinata2014}. Such systems can be realized in cold atom experiments by periodically modulating the lattice~\cite{Aidelsburger2013,Miyake2013, Baur2014,Jotzu2014, Lindner2011} to create a quasimomentum-dependent admixture of additional bands~\cite{Parker2013} or coupling between internal states~\cite{Lin2009,Cooper2011,Dalibard2011,Beri2011}. Moreover, the addition of spin-orbit coupling~\cite{Lin2011} would enable the investigation of the $\mathds{Z}_2$ invariant characterizing time-reversal invariant topological insulators~\cite{Goldman2010,Beri2011,Liu2010,Kennedy2013,Mei2012,Aidelsburger2013}.\\

\tocless{\section*{Acknowledgements}}
We acknowledge illuminating discussions with  Aris Alexandradinata, Jean-No{\"e}l Fuchs, Nathan Goldman, Daniel Greif, Lih-King Lim, Gilles Montambaux, Anatoli Polkovnikov, and Gil Refael. This work was supported by the Alfred P. Sloan Foundation, the European Commision (UQUAM, AQuS), Nanosystems Initiative Munich, the Harvard Quantum Optics Center, the Harvard-MIT CUA, NSF Grant~$\smash{\text{No. DMR-1308435}}$, the DARPA OLE program, the AFOSR Quantum Simulation MURI, the ARO-MURI on Atomtronics, and the ARO-MURI QuISM program.

%%%%%%Bibliography%%%%%%%%%
%\bibliographystyle{apsrev4-1}%apsrev4-1
%\bibliography{wilsonline_refs}
\vspace{1cm}
\tocless\input{wilsonline_main_only.bbl}
\clearpage
\renewcommand{\thefigure}{S\arabic{figure}}
\renewcommand{\theequation}{S.\arabic{equation}} 
\renewcommand{\citenumfont}[1]{S#1}
\renewcommand{\bibnumfmt}[1]{[S#1]}
\renewcommand{\thesection}{S\Roman{section}}%\arabic
\renewcommand{\thesubsubsection}{\roman{subsubsection}}

\setcounter{equation}{0}
\setcounter{figure}{0}

\tocless{\section*{Supporting Online Material}}
In these supplements, we provide theoretical background on our measurement of Wilson lines in a honeycomb lattice and additional experimental details. In \ref{sec:S:eqnmotion}, we derive the equation of motion for a single particle in a lattice in the presence of a constant force, relate it to the Wilson line, and apply our results to the honeycomb lattice. We then describe our numerical calculations of the Wilson line  in \ref{sec:S:numerical}. In \ref{sec:S:projector}, we relate the Wilson line to the cell-periodic Bloch functions of the lattice and the position operator. We then check the validity of assuming completeness of the bands in the experiment in \ref{sec:S:tbapprox}. Next, we discuss general gauge freedom in quantum mechanics and its application to our measurements in \ref{sec:S:gaugefreedom}. We then present a decomposition of the $U(2)$ Wilson line into a $U(1)$ part and an $SU(2)$ part in \ref{sec:S:u2su2}. We conclude the theoretical section of the supplements by describing the reconstruction procedure for the $SU(2)$ eigenvalues of the Wilson line in \ref{sec:S:eig_recon}. Experimental methods, including data analysis techniques, are given in \ref{sec:2}.

\tableofcontents

\section{Theoretical background}
\subsection{Dynamics in the combined lattice and gradient potential} \label{sec:S:eqnmotion}
Here, we derive the equations of motion for a particle in a lattice in the presence of a constant force $\vec{F}$. The Hamiltonian of the lattice can be written as
\begin{align}
\hat H= \sum_{\bold q,n} E_{\bold q}^n \ket{\Phi_{\bold q}^n}\bra{\Phi_{\bold q}^n},
\end{align}
where $E_{\bold q}^n$ is the energy of the $n$th band at quasimomentum $\vec{q}$ and $\ket{\Phi_{\bold q}^n}$ are the Bloch states. The Bloch states can be expressed as $\ket{\Phi_{\bold q}^n}=e^{i\vec{q}\cdot\rop} \ket{u_{\bold q}^n}$, where $\ket{u_{\bold q}^n}$ are the cell-periodic Bloch functions and $\rop$ is the position operator.

Adding a constant force $\vec{F}$ to the system results in the Schroedinger equation:
\begin{align}
i\partial_t \ket{\psi(t)}=(\hat H-\bold{F}\cdot\rop)\ket{\psi(t)},
\end{align}
where we have taken $\hbar=1.$
We assume the initial state is localized in reciprocal space at quasimomentum $\vec{q_0}$ such that 
\begin{align}
\ket{\psi(0)}=\sum_n \alpha^n(0) \ket{\Phi_{\vec{q_0}}^n}
\end{align}
where $\abs{\alpha^n(0)}^2$ gives the population in the $n$th band at time $t=0$. Making the ansatz
\begin{align}
\ket{\psi(t)}&=\sum_n \alpha^n(t) \ket{\Phi_{\bold q(t)}^n}\\
\bold q(t)&=\bold q_0+\bold F t \label{eqn:q}
\end{align}
leads to the following equations of motion for a two-band system:
\begin{align}
i\partial_t\spinor{\alpha^1(t)}{\alpha^2(t)}=\begin{pmatrix}
E^1_{\bold q(t)}-\xi^{1,1}_{\bold q(t)}& -\xi^{1,2}_{\bold q(t)} \\ 
-\xi^{2,1}_{\bold q(t)} & E^2_{\bold q(t)}-\xi^{2,2}_{\bold q(t)}
\end{pmatrix}\spinor{\alpha^1(t)}{\alpha^2(t)}
\label{eqn:1}
\end{align}
where 
\begin{align}
\xi^{n,n'}_{\bold q(t)}&=\bold A^{n,n'}_{\bold q(t)}\cdot\bold F \nonumber \\
&=i \bra{u_{\bold q(t)}^n}\partial_t \ket{u_{\bold q(t)}^{n'}}.
\end{align}
and 
\begin{align}\label{Eq:A_els}
\bold A^{n,n'}_{\bold q(t)}=i \bra{u_{\bold q}^n}\nabla_\q \ket{u_{\bold q}^{n'}}|_{\bold q=\bold q(t)}.
\end{align}
The quantity $\bold A^{n,n'}_{\bold q(t)}$ defines an intra-band ($n=n'$) and an inter-band ($n\neq n'$) Berry connection. From Eq.~\ref{eqn:1}, we see that the inter-band Berry connection drives transitions between the different bands. 

\subsubsection{The limit of infinite force}
In the limit of an infinite force, the energy terms on the diagonal are negligible compared to the geometric terms $\xi^{n,n'}_{\bold q(t)}$. In this case, Eq.~\ref{eqn:1} reduces to
\begin{align}
i\partial_t\spinor{\alpha^1(t)}{\alpha^2(t)}=\begin{pmatrix}
-\xi^{1,1}_{\bold q(t)}& -\xi^{1,2}_{\bold q(t)} \\ 
-\xi^{2,1}_{\bold q(t)} & -\xi^{2,2}_{\bold q(t)}
\end{pmatrix}\spinor{\alpha^1(t)}{\alpha^2(t)}
\end{align}
Defining $\hat{\xi}_{\q(t)}$ as the matrix with elements $\xi^{n,n'}_{\bold q(t)}$, the evolution is given by
\begin{align} 
\ket{\psi(t)}&=\mathcal{T}\text{exp}[i\int dt\hat{\xi}_{\q(t)}]\ket{\psi(0)} \nonumber \\
&\equiv\hat{\vec{W}}\ket{\psi(0)}
\end{align}
Using Eq.~\ref{eqn:q} to change the variable of integration from time to quasimomentum space recovers Eq.~2 of the main text:

\begin{align} \label{Eq:S:Wop}
\hat{\mathbf{W}}_{\mathbf{q}(0) \rightarrow \mathbf{q}(t)}=\mathcal{P}\text{exp}[i\int_\mathcal{C}d\mathbf{q} \hat{\mathbf{A}}_\mathbf{q}],
\end{align}
where the path-ordered ($\mathcal{P}$) integral runs over the path $\mathcal{C}$ in reciprocal space  from $\vec{q}(0)$ to $\mathbf{q}(t)$ and $\mathbf{\hat{A}_q}$ is the Wilczek-Zee matrix with elements defined in Eq.~\ref{Eq:A_els}. Therefore, the evolution of the system is described by an operator $\hat{\vec{W}}$, which is the Wilson line~\cite{S:Wilczek1984}. Path-ordering ($\mathcal{P}$) is necessary because the matrices generally do not commute for all quasimomenta along the path.

\subsubsection{Dynamics in the tight-binding honeycomb lattice} \label{sec:S:eqnmotion_honeycomb}

\begin{figure}[htb]
	\centering
		\includegraphics[width=80mm]{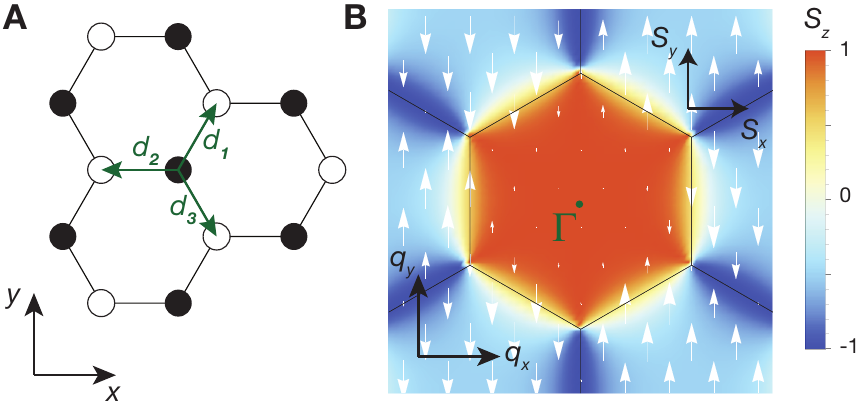}
    \caption{\textbf{The honeycomb lattice in real space and reciprocal space. A,} The real-space honeycomb lattice comprises triangular sublattices A (solid circles) and B (open circles) with nearest-neighbour hopping vectors $\vec{d_i}$. \textbf{B,} Pseudospin $\vec{S}(\q)~=~(\sin\theta_\q \cos\phi_\q,\sin\theta_\q \sin\phi_\q,\cos\theta_\q)$ for a lattice with AB-site degeneracy in a basis formed by the cell-periodic Bloch states at reference quasimomentum $\vec{Q}=\Gamma$, as labelled. The components $S_x$  and $S_y$ are indicated by white arrows, with the length of the arrow representing the magnitude of the $S_{x(y)}$ component; $S_z$ is illustrated by the color map, with red (blue) indicating $S_z>0$ $(S_z<0)$.}
	\label{Fig:S1}
\end{figure}

We now apply the results from the previous section to the specific case of the tight-binding model of the honeycomb lattice. The honeycomb lattice may be decomposed into two triangular sublattices composed of A and B sites and coupled via nearest-neighbour lattice vectors $\vec{d_i}$ with hopping amplitude~$J$ (see Fig. \ref{Fig:S1}A). We begin by defining the states $\ket{\Phi_\vec{q}^{A}}$ and $\ket{\Phi_\vec{q}^{B}}$ of the A and B sites via states $ \ket{w_{\bold r_{A (B)}}}$ localized on the $A$ ($B$) sites as
\begin{align}
\ket{\Phi_{\bold q}^A}&=\frac{1}{\sqrt{N}} \sum_{\bold r_A} e^{i \bold q \cdot \bold r_A} \ket{w_{\bold  r_A}}=e^{i\bold q \cdot \rop}\ket{u_{\bold q}^A}\\
\ket{\Phi_{\bold q}^B}&=\frac{1}{\sqrt{N}} \sum_{\bold r_B} e^{i \bold q\cdot  \bold r_B} \ket{w_{\bold r_B}}=e^{i\bold q \cdot \rop}\ket{u_{\bold q}^B},
\end{align}
where $N$ denotes the number of lattice sites. In the tight-binding limit, where the localized states $\ket{w_{\bold r_{A (B)}}}$ are eigenstates of the position operator, the cell-periodic Bloch functions $\ket{u_{\bold q}^{A (B)}}$ are independent of quasimomentum and can be physically understood as superpositions of the localized wavefunctions $\ket{w_{\bold r_{A (B)}}}$ on all A~(B) sites. 

In the basis of $\ket{\Phi_\vec{q}^{A}}$ and $\ket{\Phi_\vec{q}^{B}}$, the Hamiltonian describing the two lowest bands of the honeycomb lattice is~\cite{S:Semenoff84}
\begin{equation}
\label{eq:Htb}
\hat{H}_{\text{tb}}(\vec{q})=\begin{pmatrix}
\Delta/2 & t_\vec{q} \\
 t_\vec{q}^{*} & -\Delta/2
 \end{pmatrix},
\end{equation}
where $\Delta$ is an energy offset between the sublattices and
\begin{align}
\label{tkterm}
t_\vec{q}&=\abs{t_\q}e^{i\vartheta_\q} \nonumber\\
&=-J(e^{-i\vec{q}\cdot\dd_1}+e^{-i\vec{q}\cdot\dd_2}+e^{-i\vec{q}\cdot\dd_3}).
\end{align}
When the A and B sites are degenerate at $\Delta=0$, this Hamiltonian is diagonalized by the eigenstates
\begin{align}
\ket{\Phi_\vec{q}^1}=\frac{1}{\sqrt{2}}(\ket{\Phi_\vec{q}^{A}}-e^{i\vartheta_\vec{q}}\ket{\Phi_\vec{q}^{B}}) \\
\ket{\Phi_\vec{q}^2}=\frac{1}{\sqrt{2}}(\ket{\Phi_\vec{q}^{A}}+e^{i\vartheta_\vec{q}}\ket{\Phi_\vec{q}^{B}}).
\end{align}

The corresponding eigenenergies are
\begin{align} \label{eqn:s:energies}
E_\vec{q}^1&=-\abs{t_\vec{q}}\\ \nonumber
E_\vec{q}^2&=\abs{t_\vec{q}}.
\end{align}

\subsubsection{Elements of the Wilczek-Zee connection for $\Delta=0$}
To calculate the connections in the Wilczek-Zee matrix, we note that 
\begin{align}
\ket{u_{\bold q}^1}&=\frac{1}{\sqrt{2}}(\ket{u_{\bold q}^A}-e^{i\vartheta_{\bold q}}\ket{u_{\bold q}^B})=e^{-i\bold q\cdot \rop}\ket{\Phi_{\bold q}^1}\\
\ket{u_{\bold q}^2}&=\frac{1}{\sqrt{2}}(\ket{u_{\bold q}^A}+e^{i\vartheta_{\bold q}}\ket{u_{\bold q}^B})=e^{-i\bold q\cdot \rop}\ket{\Phi_{\bold q}^2}.
\end{align}

The intra-band (Berry) connections are then
\begin{align}\label{eqn:s:intra}
\bold A^{1,1}_{q}=\bold A^{2,2}_{q}=-\frac{1}{2}\nabla_{\bold q} \vartheta_{\bold q},
\end{align}
and the inter-band connections are
\begin{align}\label{eqn:s:inter}
\bold A^{1,2}_{q}=\bold A^{2,1}_{q}=\frac{1}{2}\nabla_{\bold q} \vartheta_{\bold q}.
\end{align}

\subsection{Numerical calculation of  Wilson lines} \label{sec:S:numerical}
Using the equations of motion derived in Eq.~\ref{eqn:1}, solving for the Wilson line transporting from quasimomentum $\q(t=0)$ to quasimomentum $\q(t=T)$ yields
\begin{align}
\hat{\vec{W}}_{\q(0)\rightarrow\q(T)}=e^{-i\int_{0}^{T}\tilde{H}(t)dt},
\end{align}
where $\tilde{H}(t)$ is the matrix on the right-hand side of Eq.~\ref{eqn:1}.

To numerically calculate the Wilson line, we compute the Trotter product of $n$ time-independent matrices evaluated at discrete time-steps of size $\Delta t=T/n$:
\begin{align}
\hat{\vec{W}}_{\q(0)\rightarrow\q(T)}&=e^{-i\int_{0}^{T}\tilde{H}(t)dt} \\ \nonumber
&\approx \prod_{j=1}^{n} e^{-i\tilde{H}(t_j)\Delta t}.
\end{align}
In our calculations, we use several hundred time-steps, depending on the length of the path.

In the tight-binding model, we have analytical expressions for the eigenenergies~(Eq.~\ref{eqn:s:energies}) and the Berry connections~(Eqs.~\ref{eqn:s:intra}~and~\ref{eqn:s:inter}); it is therefore straight-forward to calculate $\tilde{H}(t)$ for any time $t$. In the \textit{ab-initio} calculation, obtaining the Berry connections requires a particular gauge-choice to ensure that the cell-periodic Bloch functions are numerically differentiable. Similar to the approach in Ref.~\cite{S:Kohn1959}, we choose our gauge such that the Bloch functions of the $s$-bands are entirely real on a lattice site and the Bloch functions of the $p$-bands are entirely imaginary midway between neighboring lattice sites. This gauge-choice allows us to numerically differentiate the cell-periodic Bloch functions along the $q_x$ direction, which is sufficient for our experiments. 

\subsection{Wilson lines as projectors} \label{sec:S:projector}
To derive the relation between the Wilson line and the cell-periodic Bloch functions, we discretize the Wilson line in Eq.~\ref{Eq:S:Wop} by dividing the path from $\vec{q}(0)$ to $\vec{q}(t)$ into $N$ infinitesimal segments $\vec{q_1}, \vec{q_2}...\vec{q_N}.$ The Wilson line can then be expressed as a sequence of path-ordered products of projectors $\mathscr{P}(\q) = \sum_{n=1}^{\nbands}\uket{n}{\q}\ubra{n}{\q}$~\cite{S:Yu2011} with elements
\begin{equation}
\wel{mn}{q_1}{q_N}=\ubra{n}{\q_N}\prod_{i=1}^{N}\mathscr{P}(\q_i)\uket{m}{\q_1},
\end{equation}
where $\nbands$ is the number of bands. In the experiment, $\nbands=2.$

When the states $\ket{u^n_\vec{q}}$ form a complete basis over the Hilbert space $\mathcal{H}$, the projectors are trivial, i.e. $\mathscr{P}(\q) =\sum_{n=1}^{\nbands}\ket{u^n_{\q}}\bra{u^n_{\q}} = \mathds{1}$. In this case, the Wilson line elements reduce to the overlap of the cell-periodic Bloch functions:
\begin{equation}\label{eq:S:WilsonProj}
\wel{mn}{q_1}{q_N}=\braket{u^n_{\q_N}}{u^m_{\q_1}}.
\end{equation}
Therefore, the Wilson line provides a way of comparing the cell-periodic Bloch states $\uket{n}{\q}$ at any two points in momentum space. In terms of the Bloch states $\ket{\Phi_{\bold q}^n}=e^{i\vec{q}\cdot\rop} \ket{u_{\bold q}^n}$, the Wilson line elements can equivalently be expressed as
\begin{align}\label{Eq:realspace_wel}
\wel{mn}{q_1}{q_N} &=\bra{\Phi^n_{\q_N}}e^{i\q_N\cdot\rop}e^{-i\q_1\cdot\rop}\ket{\Phi^m_{\q_1}}\nonumber\\
&=\bra{\Phi^n_{\q_N}}e^{i\Delta\q\cdot\rop}\ket{\Phi^m_{\q_1}}
\end{align}
where $\Delta\q=\q_N-\q_1$ is the change in quasimomentum. Therefore, expressed in terms of the real-space position operator $\rop$, the Wilson line transporting a state from an initial quasimomentum $\q_i$ to final quasimomentum $\q_f$ by $\Delta\q$ is 
\begin{align}
\hat{\vec{W}}_{\q_i\rightarrow\q_f}=e^{i\Delta\q\cdot\rop}.
\end{align}

\subsection{Validity of the two-band tight-binding approximation} \label{sec:S:tbapprox}
In this section, we ascertain the validity of applying Eq.~\ref{eq:S:WilsonProj} to our system, which only holds when the two-lowest band eigenstates span the same Hilbert space at all quasimomenta.  To this end, we use the numerically-calculated band eigenstates of the full optical lattice potential to calculate the elements of the Wilczek-Zee connection $\bold A^{n,n'}_{q}$ for $n,n'=1,2$. We compare the corresponding Wilson lines at various lattice depths to the Wilson line obtained from a tight-binding calculation.

\begin{figure}[htb]
	\centering
		\includegraphics[width=80mm]{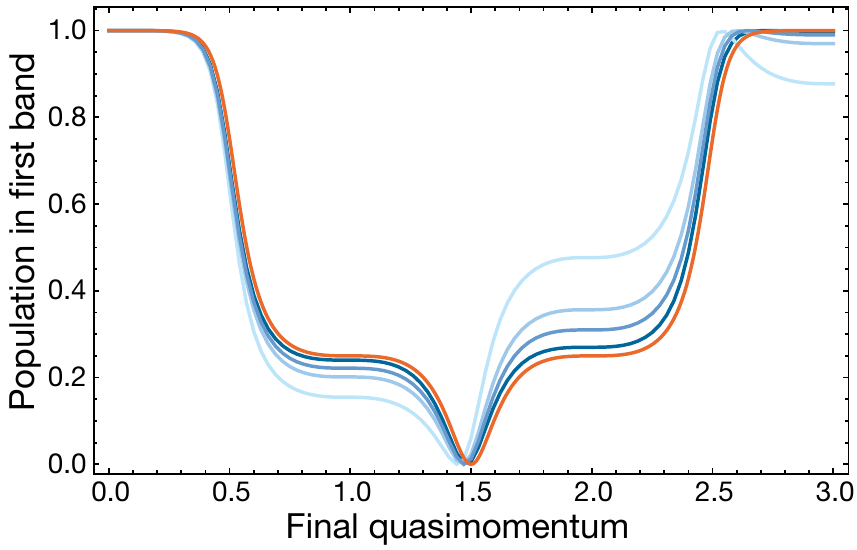}
    \caption{\textbf{Validity of the two-band tight-binding approximation.} Lowest band population, $\abs{W^{11}_{\Gamma\rightarrow\q}}^2$ for a two-band Wilson line using  a full band structure calculation for lattice depths 5.2$E_r$, 3$E_r$, 2$E_r$, and 1$E_r$ (decreasing lightness of blue). The Wilson line at $5.2E_r$ is nearly identical to the Wilson line obtained from a tight-binding calculation (orange).}
	\label{Fig:Stwobandcomp}
\end{figure}

We plot the population in the first band, $\abs{W^{11}_{\Gamma\rightarrow\q}}^2$ at different lattice depths for a path from $\Gamma$ to $\Gamma+3\vec{G}$ in Fig.~\ref{Fig:Stwobandcomp}.  While strong deviations occur for shallow lattices,  only minor differences are visible at the lattice depth of $5.2E_r$ used in the experiment. Therefore, the experiment should, in principle, be well-described by the tight-binding formalism.  

However, in Fig.~3 of the main text, the population in the lowest band returns to only 90$\%$ and not unity at $\Gamma+3\vec{G}$. This is a result of transfer into higher bands. Due to the presence of higher bands, we can not exactly realize the two-band Wilson lines plotted in  Fig.~\ref{Fig:Stwobandcomp}, which would require reaching the infinite gradient limit for the two lowest bands while remaining adiabatic with respect to higher bands. Experimentally, the choice of gradient strength is a compromise between realizing dynamics that are fast compared to the energy scale of the lowest two bands and minimizing excitations into higher bands. We could, in principle, realize the two-band Wilson line regime more precisely by increasing the lattice depth, which decreases the combined width $\varepsilon$ of the lowest two bands and increases the energy scale between the lowest two bands and higher bands. However, in the current work, the lattice depth was limited by the bandmapping technique. As the lattice depth is increased, it becomes more difficult for the bandmapping process to remain adiabatic with respect to $\varepsilon$. 

\subsection{Gauge Freedom in Wilson lines}  \label{sec:S:gaugefreedom}
Due to its linear structure, quantum mechanics contains an inherent gauge freedom: if $\ket{\psi}$ is an eigenstate of an operator, then so is $e^{i\phi}\ket{\psi}$, $\phi\in\mathbb{R}$. Consequently, even non-degenerate eigenstates are defined uniquely only up to such a phase factor. If the Hamiltonian depends on a parameter $\vec{R}$ as $\hat{H}(\vec{R})$, then the corresponding phase $\phi(\vec{R})$ can be chosen independently for every value of $\vec{R}$. This is called a local $U(1)$ gauge freedom, in reminiscence of the situation in quantum electrodynamics or gauge theories~\cite{S:Peskin1995}.

Due to this gauge freedom, the Berry phase for adiabatic evolution in a non-degenerate band is only well-defined for closed loops, since, for non-closed paths, the phase factor in front of the final state depends on the choice of basis, i.e., the gauge choice, at the final point.
In the situation of two, everywhere non-degenerate eigenstates \ket{1} and \ket{2}, the gauge group enlarges to $U(1)\times U(1)$, since the phase of each eigenstate can be chosen independently.
However, if the two eigenstates are degenerate with eigenenergy $E$, then any normalized superposition $\ket{\Psi}=\alpha\ket{1}+\beta\ket{2}$ is also an eigenstate with the same eigenvalue:
\begin{align} 
\hat{H}\ket{\Psi}&=\hat{H}\left(\alpha\ket{1}+\beta\ket{2}\right) \nonumber \\
&=\alpha\hat{H}\ket{1}+\beta\hat{H}\ket{2} \nonumber \\
&=E\ket{\Psi} 
\end{align} 
The freedom in choosing the basis states is therefore now enlarged. For each normalized superposition state \ket{\Psi}, there exists a corresponding orthogonal superposition state \ket{\Psi_\perp} such that \ket{\Psi} and \ket{\Psi_\perp} form an orthonormal basis. Consequently, the gauge freedom is U(2). 

The Hamiltonian is, however, not the only observable. If there exists another observable that is conserved by all $\hat{H}(\vec{R})$ and can distinguish between the states \ket{1} and \ket{2}, such as spin or parity, then this observable defines a new, more constrained basis. Assuming the basis states to be eigenstates of both the Hamiltonian and the observable, the gauge freedom is reduced again to $U(1)\times U(1)$, even for degenerate eigenenergies. This is precisely the case in the strong-force limit of the experiment: even though the bands appear degenerate during the evolution, i.e., the difference in eigenenergies is negligible during the dynamics, the bandmapping procedure can nonetheless distinguish between the two bands.  In such a basis, the absolute values of the Wilson line elements are well-defined and can be observed even for open lines.

\subsubsection{The pseudospin representation and the Wilson line elements}
We now explicitly discuss the gauge-invariance of the terms  $\theta_\vec{q}$ and  $\phi_\vec{q}$ used in the pseudospin representation of the cell-periodic Bloch state. The components of the pseudospin referenced to $\vec{Q}=\Gamma$ is plotted in Fig.~\ref{Fig:S1}B for a lattice with AB-site degeneracy.

The polar angle $\theta_\vec{q}$ is obtained from the quantity
\begin{align}
\abs{\braket{u_\vec{q}^1}{1}}&=\abs{\cos\frac{\theta_\vec{q}}{2}} \nonumber \\
\Rightarrow \theta_\vec{q}&=2\text{arccos}\abs{\braket{u_\vec{q}^1}{1}} \nonumber \\
&=2\text{arccos}\abs{\wel{11}{\vec{Q}}{\vec{q}}}
\end{align}
where we have used Eq. \ref{eq:S:WilsonProj} to relate the overlap of the cell-periodic Bloch functions to the Wilson line elements in the last line. The $U(1)\times U(1)$ gauge-freedom on the phase of $\ket{u^1_\vec{q}}$ or the reference states $\ket{1}$ and $\ket{2}$ does not affect the absolute value of the overlap. Hence, $\abs{\wel{11}{\vec{Q}}{\vec{q}}}$ and, consequently, $\theta_\vec{q}$ is a gauge-invariant quantity.

The relative phase $\phi_\vec{q}$ can be expressed as
\begin{align}
\phi_\q&=\text{Arg}[\braket{u_\vec{q}^1}{1}]-\text{Arg}[\braket{u_\vec{q}^1}{2}] \nonumber\\
&=\text{Arg}[\wel{11}{\vec{Q}}{\vec{q}}]-\text{Arg}[\wel{12}{\vec{Q}}{\vec{q}}] 
\end{align}
That is, we access the relative phase between the basis states by obtaining the difference of phases between the Wilson line elements. While the gauge-freedom of $\ket{u_\vec{q}^1}$ is cancelled out by taking the difference of the argument of Wilson line elements, the gauge-freedom on the reference states remains. A different choice of the phase for $\ket{1}$ and $\ket{2}$ changes the value of $\phi_\vec{q}$. However, to obtain a gauge-invariant quantity, we can compare $\phi_\vec{q}$ and $\phi_\vec{q'}$ at quasimomenta $\vec{q}$ and $\vec{q'}$. Taking the difference between the relative phases at the two quasimomenta cancels out the gauge-freedom of the reference states if the same reference states are used to define the cell-periodic Bloch function at both quasimomenta. Explicitly, the gauge-invariant quantity measured in the experiment is
\begin{align} \label{Eq:inv_phase}
\phi_\vec{q}-\phi_\vec{q'},
\end{align}

\subsection{Decomposition of the Wilson line into U(1) and SU(2)} \label{sec:S:u2su2}
Here, we decompose the U(2) Wilson line into an SU(2) matrix multiplied by a global U(1) phase. We furthermore relate the U(1) phase to the sum of the Berry phase of the first and second band, noting, however, that the U(1) part might not show up in certain bases~\cite{S:Zee1988}, and that our experiment measures only the SU(2) part of the Wilson line.

We begin by writing the Wilczek-Zee connection matrix $\hat{\vec{A}}_\vec{q}$ in a general form as
\begin{align}
\hat{\bold  A}_\vec{q}=
\begin{pmatrix}
\bold A^{1,1}_\vec{q}& \bold A^{1,2}_\vec{q} \\ 
\bold A^{2,1}_\vec{q} & \bold A^{2,2}_\vec{q}
\end{pmatrix}
\end{align}
To simplify notation, we henceforth suppress the $\vec{q}$ subscript and note that all elements are still to be understood as being $\vec{q}$-dependent. 
Next, we decompose the matrix as
\begin{align}
\hat{\bold  A}&=
\begin{pmatrix}
\frac{\bold A^{1,1}+\bold A^{2,2}}{2}& 0 \\ 
0 & \frac{\bold A^{1,1}+\bold A^{2,2}}{2}
\end{pmatrix}+
\begin{pmatrix}
\frac{\bold A^{1,1}-\bold A^{2,2}}{2}& \bold A^{1,2} \\ \notag
 \bold A^{2,1}& -\frac{\bold A^{1,1}-\bold A^{2,2}}{2}
\end{pmatrix}\\
&:= \hat{\bold A}_{U(1)} + \hat{\bold A}_{SU(2)}
\end{align}

Noting that $\hat{\bold A}_{U(1)}$ is proportional to the identity matrix and therefore commutes with $\hat{\bold A}_{SU(2)}$ and with itself for all quasimomenta, the Wilson line $\hat{\vec{W}}_{\vec{Q}\rightarrow\vec{q}}$ transporting a state from initial quasimomentum $\vec{Q}$ to final quasimomentum $\vec{q}$ can be expressed as:
\begin{align}
\hat{\vec{W}}_{\vec{Q}\rightarrow\vec{q}}&=\mathcal{P}e^{i\int_\mathcal{C}d\vec{q} \hat{\bold A}} \nonumber\\
&=\mathcal{P}e^{i\int_\mathcal{C}d\vec{q} \hat{\bold A}_{U(1)}+ \hat{\bold A}_{SU(2)}} \nonumber\\
&=e^{i\int_\mathcal{C}d\vec{q} \hat{\bold A}_{U(1)}}\mathcal{P}e^{i\int_\mathcal{C}d\vec{q}\hat{\bold A}_{SU(2)}}
\end{align} 
where $\mathcal{C}$ denotes the path taken from $\vec{Q}$ to $\vec{q}$. This gives the decomposition of the $U(2)$ Wilson line into a $U(1)$ global phase multiplied by a path-ordered $SU(2)$ matrix. Furthermore, the global $U(1)$ phase is given by
\begin{align}
\int_C d\vec{q}\frac{\bold A^{1,1}+\bold A^{2,2}}{2} = \frac{\phi_1+\phi_2}{2}
\end{align}
where $\phi_1$ is the adiabatic phase acquired in the first band and $\phi_2$ is the adiabatic phase acquired in the second band. For closed loops, $\phi_1$ and $\phi_2$ are the Berry phases, which can be used to formulate the Chern number of the system~\cite{S:Thouless1982}. For paths closed only by a reciprocal lattice vector, $\phi_1$ and $\phi_2$ are instead the Zak phases~\cite{S:Zak1989}.

\subsection{Reconstruction of SU(2) Wilson-Zak loops} \label{sec:S:eig_recon}
Generically, an SU(2) matrix can be expressed as
\begin{align}
\begin{pmatrix}
W^{11} & W^{12}\\
-W^{12*} & W^{11*}
\end{pmatrix}
\end{align}
where $\abs{W^{11}}^2+\abs{W^{12}}^2=1$. The eigenvalues $e^{\pm i\xi}$ of this matrix depend on the absolute values of the Wilson line terms and the phase of $W^{11}$ and are given by
\begin{align}
\text{Re}[{W^{11}}]\pm i\sqrt{\abs{W^{12}}^2+\text{Im}[W^{11}]^2}
\end{align}

The absolute values are directly measured via the population remaining in the first band after transport of the lowest band eigenstate. Although the interferometric sequence reveals only the difference between the phases of elements $W^{11}$ and $W^{12}$, we can extract the phase of $W^{11}$ by invoking unitarity and the "backtracking" condition of Wilson lines. It can be shown that, for generic quasimomenta $\vec{q}$ and $\vec{Q}$, $\hat{\vec{W}}_{\vec{Q}\rightarrow\vec{q}}=\hat{\vec{W}}^\dagger_{\vec{q}\rightarrow\vec{Q}}$~\cite{S:Makeenko2005}. Consequently, $\hat{\vec{W}}_{\vec{q}\rightarrow\vec{Q}}\hat{\vec{W}}_{\vec{Q}\rightarrow\vec{q}}=\mathds{1}$,  such that going forward and back along the same path results in no transformation of the state vector.

In the case of the Wilson-Zak loops in the experiment, the relevant relation is
\begin{align} \hat{\vec{W}}_{\Gamma\rightarrow\Gamma+\vec{G}}=\hat{\vec{W}}^\dag_{\Gamma\rightarrow\Gamma-\vec{G}}
\end{align} 
where we have used that $\hat{\vec{W}}_{\Gamma\rightarrow\Gamma-\vec{G}}=\hat{\vec{W}}_{\Gamma+\vec{G}\rightarrow\Gamma}$ which, assuming a periodic gauge choice, follows from the periodicity of the BZ. If the phase $\phi$ of the oscillation after transport from $\Gamma$ to $\q_\alpha=\Gamma+\vec{G}$ is given by 
\begin{align}
\phi=\phi_p+\text{Arg}[W^{11}_{\Gamma \rightarrow \mathbf{q_\alpha}}]-\text{Arg}[W^{12}_{\Gamma \rightarrow \mathbf{q_\alpha}}]
\label{eqn:S:phase}
\end{align}
where $\phi_p$ is a gauge-dependent quantity that results from the initial transport to prepare the superposition state, 
then the phase $\phi'$ of the oscillation after transport from $\Gamma$ to $\Gamma-\vec{G}$ is 
\begin{align}\label{eqn:S:phasep}
\phi'=\phi_p-\text{Arg}[W^{11}_{\Gamma \rightarrow \Gamma+\vec{G}}]-\text{Arg}[W^{12}_{\Gamma \rightarrow \Gamma+\vec{G}}] + \pi
\end{align}
Therefore, taking the difference between the two oscillation phases extracts $\text{Arg}[W^{11}_{\Gamma \rightarrow \Gamma+\vec{G}}]$ as
\begin{align}
\phi-\phi'-\pi=2\text{Arg}[W^{11}_{\Gamma \rightarrow \Gamma+\vec{G}}]
\label{phasediff}
\end{align}
Note that there is an ambiguity in choosing $\pm\pi$ when relating $\phi'$ to $\phi$. This results in a global $U(1)$ phase shift of $\pi$ in the eigenvalue phases. However, the phase difference between the eigenvalues is unaffected, which is sufficient to reconstruct, e.g., the $\mathds{Z}_2$ invariant.

When the A and B sites of the lattice are degenerate, applying Eq.\ref{phasediff} to the phase of oscillations for $\smash{\alpha~=~0}$ and $\smash{\alpha~=~180}$ in the interferometric sequence (Fig.~4B of main text) yields Arg$[W^{11}_{\Gamma\rightarrow\Gamma+\vec{G}}]$=0.03(7) rad. Combined with the direct transport data from $\Gamma$ to $\Gamma+\vec{G}$ (Fig.~3B of main text), which gives $\abs{W^{11}_{\Gamma\rightarrow\Gamma+\vec{G}}}$=0.47(2) and $\abs{W^{12}_{\Gamma\rightarrow\Gamma+\vec{G}}}=\sqrt{1-\abs{W^{11}_{\Gamma\rightarrow\Gamma+\vec{G}}}^2}$=0.88(1), we obtain eigenvalues exp$[\pm i1.03(2)\pi/3]$.

\begin{figure}[htb]
	\centering
		\includegraphics[width=80mm]{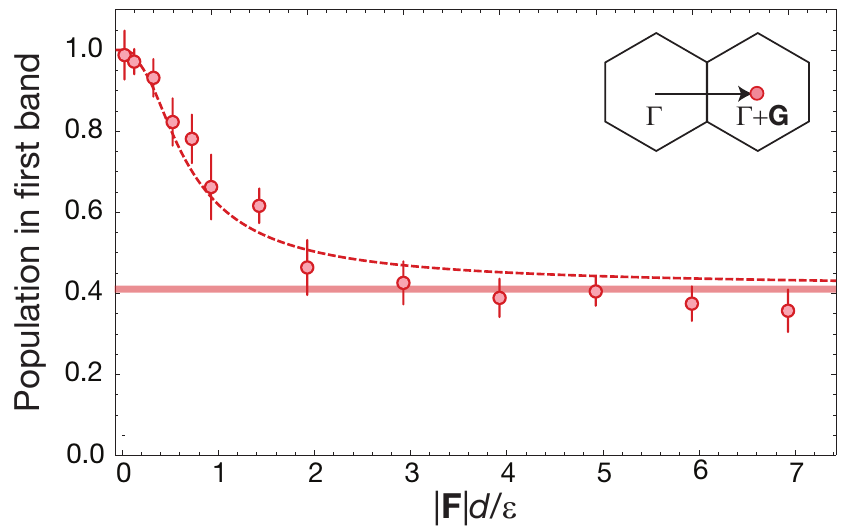}
    \caption{\textbf{Reaching the Wilson line regime in a lattice with AB-site offset.} The population remaining in the first band after transport at different forces from $\Gamma$ to $\Gamma+\vec{G}$. The data agrees reasonably well with a two-level, tight-binding theory (dashed line) that approaches the Wilson line regime (thick shaded line) at large forces. We attribute the discrepancy to the two-level model at larger forces to transfer to higher bands. To calculate the SU(2) eigenvalues, we use the population at $\abs{\vec{F}}d/\varepsilon=5$. The inset depicts the transport path.  Error bars represent the standard error of the mean from ten shots per data point. }
	\label{Fig:S2}
\end{figure}

Similarly, in the lattice with AB-site offset, data for $\smash{\alpha~=~0}$ and $\smash{\alpha~=~180}$ from Fig.~4B in the main text yields Arg$[W^{11}_{\Gamma\rightarrow\Gamma+\vec{G}}]$=-0.76(6) rad. We measure the absolute values by transporting atoms initialized at $\Gamma$ in the lowest eigenstate to $\Gamma+\vec{G}$ with increasing force. The remaining population in the first band is shown in Fig.~\ref{Fig:S2}. The eventual saturation of population transfer indicates the geometric nature of the transfer. At $\abs{\vec{F}}d/\varepsilon=5$, we obtain $\abs{W^{11}_{\Gamma\rightarrow\Gamma+\vec{G}}}$=0.63(3) and $\abs{W^{12}_{\Gamma\rightarrow\Gamma+\vec{G}}}=\sqrt{1-\abs{W^{11}_{\Gamma\rightarrow\Gamma+\vec{G}}}^2}$=0.77(2). The eigenvalues of this Wilson-Zak loop are then exp$[\pm i 1.04(4)\pi/3]$.

\section{Experimental methods} \label{sec:2}
\subsection{The optical potential of the honeycomb lattice} \label{sec:S:opticalpotential}
The total potential resulting from interfering three beams of variable polarization at 120$^\circ$ angles can be decomposed into the sum of its out-of-plane ($s$-) and in-plane ($p$-) components as
\begin{align}
V(x,y)=&V^s(x,y)+V^p(x,y)\nonumber \\
=&\abs{\sum^3_{i=1}\sqrt{V_i^s}e^{-i\k_i\cdot\mathbf{r}}}^2+\abs{\sum^3_{i=1}\sqrt{V_i^p}e^{-i(\k_i\cdot\mathbf{r}-\alpha_i)}}^2
\end{align}
where $V^{s(p)}_i$ is the ac Stark shift produced by the $s(p)$-component, $\k_i$ is the wave-vector with wavenumber \smash{$k_{L}=\abs{k_i}$}, and $\alpha_i$ is the phase between $s$- and $p$-polarization components of beam $i$. 

In the lattice with AB-site degeneracy, all three beams have equal intensity and are purely $s$-polarized. In this case, the expression for the total potential reduces to
\begin{align}
V(x,y)=V_0\Big(&2\cos(\sqrt{3}k_L x) \nonumber \\
&+ 4\cos(\frac{\sqrt{3}k_L x}{2})\cos(\frac{3k_Ly}{2})+3\Big) 
\end{align}
where $V_0\equiv V^s_1=V^s_2=V^s_3$.

The resulting honeycomb potential contains two non-equivalent lattice sites (A,B) per unit cell, as shown in Fig. \ref{Fig:S3}. Consequently, the two lowest bands, which correspond to the $\textit{s}$-orbitals on the A and B sites, touch at Dirac points and are strongly coupled to each other. Coupling to the next higher bands, however, can mostly be neglected due to the large energy gap to the $\textit{p}$-orbitals. Hence, our experimental system is well-approximated by a two-band model.

\subsubsection{Breaking AB-site degeneracy}
\begin{figure*}[htb]
	\centering
		\includegraphics[width=160mm]{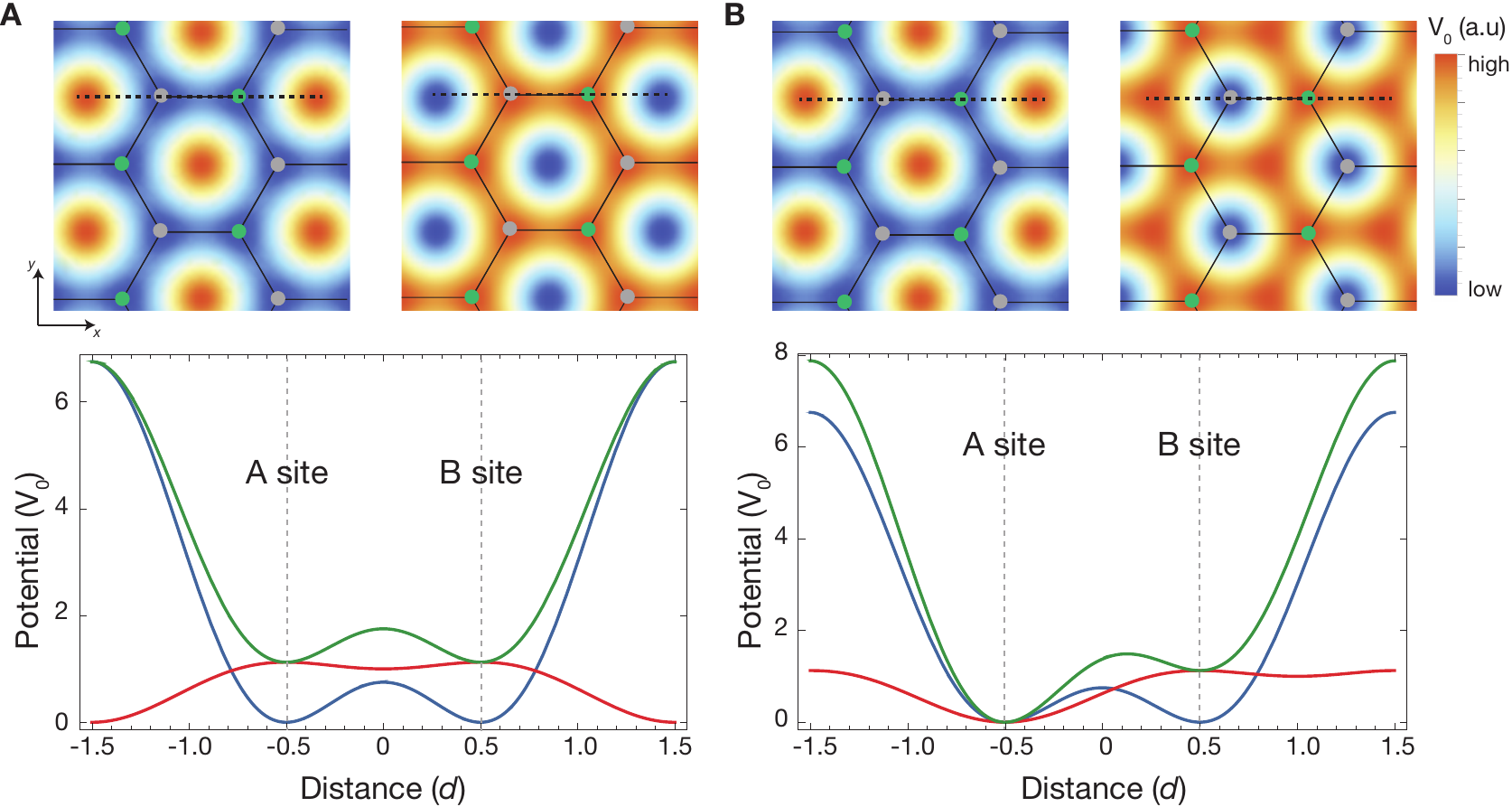}
    \caption{\textbf{The honeycomb lattice potential. a,} A lattice with degenerate A and B sites formed by beams of polarization angle $\theta=\pi/6$ and $\alpha_{32}=\alpha_{13}=0$. Top: 2D plots of the $s$-polarized (left) and $p$-polarized (right) components of the potential.  In the $s(p)$-polarized potential, the A (gray circles) and B sites (green circles) are both located at the potential minima (maxima). Therefore, there is no energy offset in the total potential, which is the sum of the two polarization components at the A and B sites. Bottom: A cross-cut of the potential through the dashed line in the 2D plots. The $s(p)$-potential is in blue (red) and the total potential is in green. \textbf{b,} Same as in (\textbf{a}), but for $\alpha_{32}=2\pi/3$ and $\alpha_{13}=2\pi/3$. With the appropriate phase shift between the polarization components of the beams, the A sites are located at the minima while the B sites are located at the maxima of the $p$-polarized component of the potential. Consequently, there is an energy offset between A and B sites in the total potential.}
	\label{Fig:S3}
\end{figure*}

To introduce an energy-offset between the A and B sites and maintain isotropic tunnelling, we constrain the polarizations of each beam to have the same composition of $s$- and $p$-polarizations. The potentials arising from the interference of the $s$- and $p$- components of the three beams, which are shown separately in Fig. \ref{Fig:S3}, have the form
\begin{align}
V^s(x, y)=V_0\cos^2\theta \Bigg(&2\cos(\sqrt{3}k_L x) \nonumber \\
&+ 4\cos(\frac{\sqrt{3}k_L x}{2})\cos(\frac{3k_Ly}{2})+3\Bigg) 
\end{align}
and
\begin{align}
V^p(x, y)= &-V_0\sin^2\theta\Bigg(\cos(\frac{\sqrt{3}k_Lx}{2} + \frac{3k_Ly}{2} - \alpha_{32}) \nonumber\\
&+\cos(\frac{\sqrt{3}k_Lx}{2} - \frac{3k_Ly}{2} + \alpha_{13} + \alpha_{32})\nonumber\\ 
&+\cos(\sqrt{3}k_Lx + \alpha_{13})-3\Bigg) 
\end{align}
where $\alpha_{32}\equiv\alpha_3-\alpha_2$ and $\alpha_{13}\equiv\alpha_1-\alpha_3$ and $\theta$ parametrizes the composition of $s$- and $p$-polarizations, i.e., for $\theta$=0, the light is purely $s$-polarized and for $\theta=\pi/2$, the light is purely $p$-polarized. Furthermore, in defining the same $V_0$ for the $s$- and $p$-polarizations, we have neglected the state-dependence of the dipole potential, which is valid in our case of far-detuned light.

By choosing the phase $\alpha_i$ of each beam, we can shift the $p$-polarized potential relative to the $s$-polarized potential. When $\alpha_{32}=\alpha_{13}=0$, the minima of the $p$-polarized potential and the maximima of the $s$-polarized potential coincide with the A and B sites (Fig. \ref{Fig:S3}A). Subsequently, atoms experience the same ac Stark shift in either an A or B site. However, by setting $\alpha_{32}=2\pi/3$ and $\alpha_{13}=2\pi/3$, the $s$-polarized potential is shifted such that the potential maxima occur on A sites while the potential minima occur on B-sites (Fig. \ref{Fig:S3}B). 

\subsubsection{Implementation of the honeycomb lattice}
For the lattice with AB-site degeneracy, the polarization of the three beams is set by polarizing beam splitters. We have verified in previous work \cite{S:Duca2015} that this results in sufficiently pure $s$-polarizations.

To introduce an AB-site offset, we tune the polarizations of the beams by using a half- and a quarter-waveplate in the paths of two beams and only a half-waveplate in the path of the third beam, which does not require a phase shift between its $s$- and $p$-polarized components. After setting the waveplates, we ensure that the polarization composition of each beam is approximately equal by taking time-of-flight (TOF) images of the BEC after sudden release from the lattice. An unequal polarization composition between the beams results in an imbalance in the Bragg peaks. We then check the dispersion relation through the Ramsey-like interferometric procedure described in Sec.~\ref{sec:S:eig_recon}. To quantitatively assess the amount of AB-site offset, we fit the measured dispersion relation to a tight-binding model~(see Fig.~\ref{Fig:S6}).

\subsection{Preparation scheme}\label{sec:atomprep}
The evaporative cooling of $^{87}$Rb atoms in the $\smash{\ket{F=1,~m_F=1}}$ state to quantum degeneracy is initiated in a plugged quadrupole trap and completed in a crossed-beam dipole trap of wavelength 1064 nm. At the end of the cooling process, we have approximately 4$\times 10^4$ atoms in the BEC. The atoms are adiabatically loaded into a honeycomb optical lattice of depth $5.2(1)E_r$ in 100 ms. During the experimental sequence, the combined trap frequencies of the blue-detuned lattice and dipole potential are $\omega_z=118(9)$~Hz and $\omega_{xy}=16(1)$~Hz. We obtain these frequencies by measuring the oscillation frequency of the center-of-mass motion of the BEC after a perturbation of the trapping potential. We neglect the effect of the dipole trap since the dynamics of our experiment (on the order of 500~$\mu$s) is much shorter than the inverse dipole trap frequencies. 

\subsection{Lattice acceleration} 
To transport the atoms in reciprocal space, we generate a constant inertial force in the lattice frame by uniformly accelerating the lattice.
An acceleration of $\mathbf{a}_i=\frac{2}{3}\lambda_L\frac{d\nu}{dt}\hat{\mathbf{e}}_i$ in the propagation direction $\hat{\mathbf{e}}_i$  of beam $i$ is accomplished via a linear sweep of the frequency shift $\nu_i = \delta\omega_i / (2\pi)$ at a rate $\frac{d\nu_i}{dt}$. Individual control over the frequency sweep rate of two beams enables lattice acceleration of variable magnitude and direction. Thus, we can move the atoms along arbitrary paths in reciprocal space.
\begin{figure}[htb]
	\centering
		\includegraphics[width=80mm]{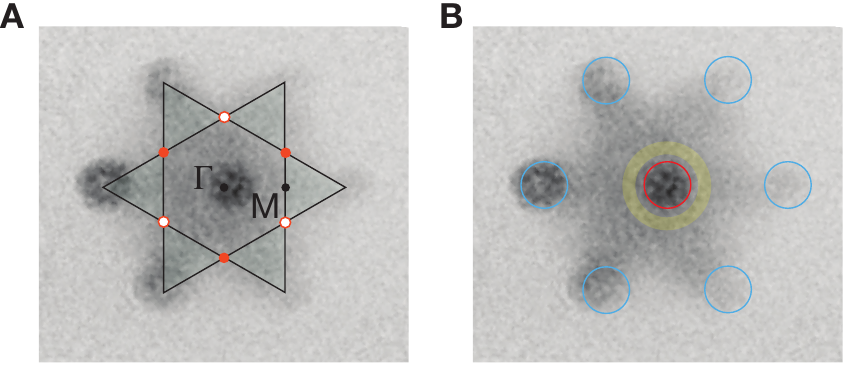}
    \caption{\textbf{Raw data of band mapped atoms at $\Gamma$. a,} The extended zone scheme showing the first (hexagon) and second (triangles) BZs is overlaid on a raw image of the band mapped atoms at $\Gamma$. High-symmetry points $\Gamma$, at the center of the first BZ, and M, at the edge of the first BZ, are labelled. Non-equivalent Dirac points $\KK$ ($\KKp$) are depicted by solid (open) orange circles at the corners of the first BZ.  \textbf{b,} Analysis ROIs. The atom number in the first (second) zone is obtained by summing the pixel values within the red (blue) circle(s). We additionally take the mean of the pixel values in the yellow ring and, with the exception of the interferometric data, subtract this value as background from the pixel sum of the first zone atoms. }
	\label{Fig:S4}
\end{figure}

\subsection{Detection} 
The detection procedure begins with a linear ramp-down of the lattice intensity in 800 $\mu$s to band map the atoms. We then use absorption imaging to detect the atoms after 9 ms TOF. Due to the short TOF, the resulting image is a convolution of the insitu cloud size and the quasimomentum. Nonetheless, for quasimomenta near the center of the first BZ, we can easily distinguish between atoms in the first and second BZs, as shown in Fig. \ref{Fig:S4}A. In contrast, it is difficult to differentiate between first and second zone atoms at the edges of the BZ. For these quasimomenta, we add an additional adiabatic segment to the sequence to push the atoms away from the edge, toward the center of the BZ, before bandmapping.

\subsubsection{Data Analysis} 
We sum the pixel values of the atoms in the first zone $n_1$ and the pixel values of the atoms in the second zone $n_2$ to obtain the fraction of atoms in the lowest band, $n_1/(n_1+n_2)$. Since we wish to count atoms localized at specific quasimomenta, we specify regions of interest (ROIs), which are depicted in Fig.\ref{Fig:S4}B for atoms at $\Gamma$. We sum pixel values within the red circle to obtain $n_1$ and sum the pixel values within the six blue circles to obtain $n_2$.

For a quantitatively accurate fraction in the lowest band, we subtract the mean pixel value of the shaded yellow region from $n_1$ to account for the hot background atoms. We do not subtract an additional background for $n_2$ since atoms in the upper band at $\Gamma$ are unstable due to interaction effects  and move to other quasimomenta. An additional background subtraction would therefore underestimate atoms in the second band by counting atoms that have decayed from $\Gamma$ as background. This analysis method was used to obtain the population in the lowest band shown in Figs.~2 and 3 in the main text. 

In contrast, for the oscillation data in Fig.~4B of the main text, we do not perform a background subtraction for $n_1$. Due to the long hold times, the cloud heats and disperses in reciprocal space, increasing the background value. Therefore, background subtraction would lead to an underestimation of atoms in the first zone. However, an overall offset in the first zone population does not affect the phase of the oscillation, which is the relevant quantity. 

To check systematic errors due to our selection of ROIs, we analyse a single dataset of population transfer vs. force magnitude after transport by one reciprocal lattice vector (blue data in Fig.~2B of main text) using different ROIs. We evaluate both the effect of the ROI size using a fixed background subtraction ring and the effect of the background subtraction ring using a fixed ROI size. Using the same ROI size for first and second band atoms and restricting the ROI size such that it does not overlap with the background subtraction ring yields consistent results with deviations on the order of $\pm5\%$. 

We use the same ROIs and, when applicable, background subtraction ring for all datasets. 

\begin{figure*}[!t]
	\centering
		\includegraphics[width=160mm]{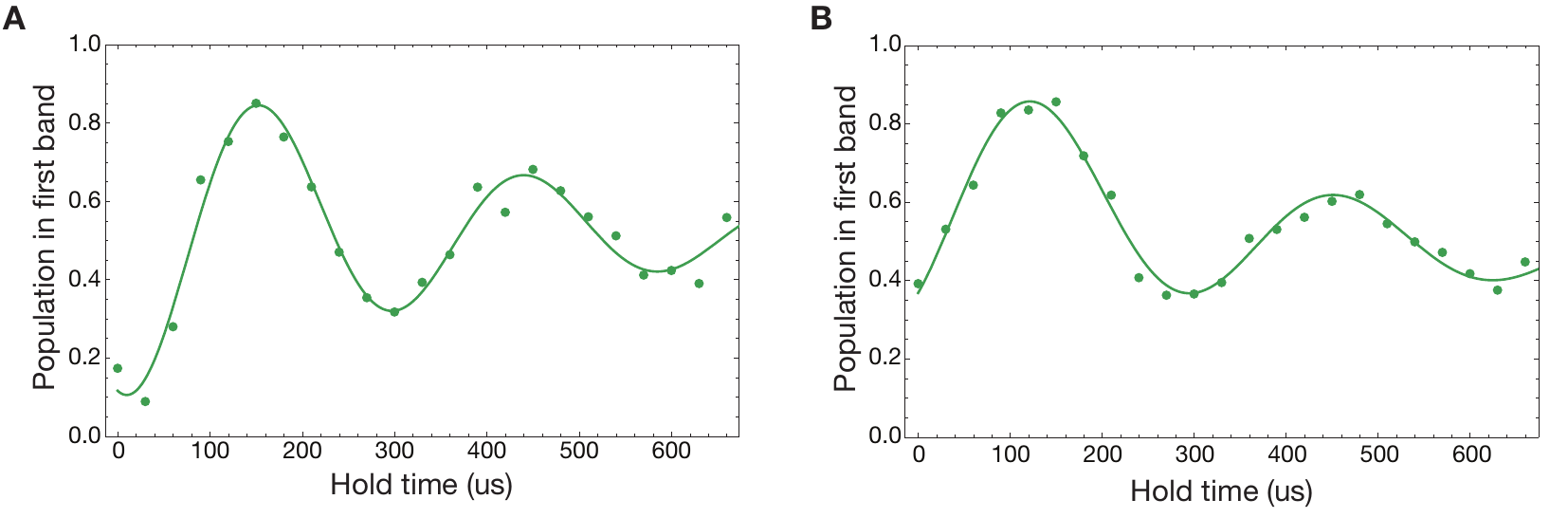}
    \caption{\textbf{Example oscillations from the Ramsey-like interferometric sequence.} Both the maximum and minimum values of the interference fringe for $\alpha=120^\circ$ in a lattice with AB-site offset (\textbf{a}) damps with increasing hold time, in contrast to the interference fringe for $\alpha=120^\circ$ in a lattice with AB-site degeneracy (\textbf{b}). Here, only the maximum values damp with increasing hold time.}
	\label{Fig:S5}
\end{figure*}

\subsection{Fitting the interference fringe}

The population in the first band $P_1(t)$ resulting from the Ramsey-like interferometric sequence described in the main text oscillates as a function of hold time and is given by:
\begin{align}
P_1(t)=C_0+A_0\text{cos}(\varepsilon t+\phi)
\end{align} 
where $C_0$ is a constant offset, $A_0$ parametrizes the amplitude of the oscillation, and the phase $\phi$ is given by Eq.~\ref{eqn:S:phase}.

To extract the phase of the interference fringe, we fit the population of the lowest band at quasimomentum $\mathbf{q_\alpha}$ to an empirically chosen function of the form:
\begin{align}
A_0e^{-t/t_0}(\cos(2\pi ft+\phi)+y_1)+y_0
\end{align}
where $A_0$ is the amplitude of the function, $t_0$ parametrizes the decay of the fringe, $f$ gives the frequency, which is determined by the dispersion at the reference quasimomentum, and $\phi$ is the phase. The offsets $y_1$ and $y_0$ interpolate between an oscillation with damping of both maximum and minimum values (Fig. \ref{Fig:S5}A) and an oscillation with damping of only the maximum values (Fig.\ref{Fig:S5}B). 

\subsection{Accessing the dispersion relation}

\begin{figure}[htb]
	\centering
		\includegraphics[width=80mm]{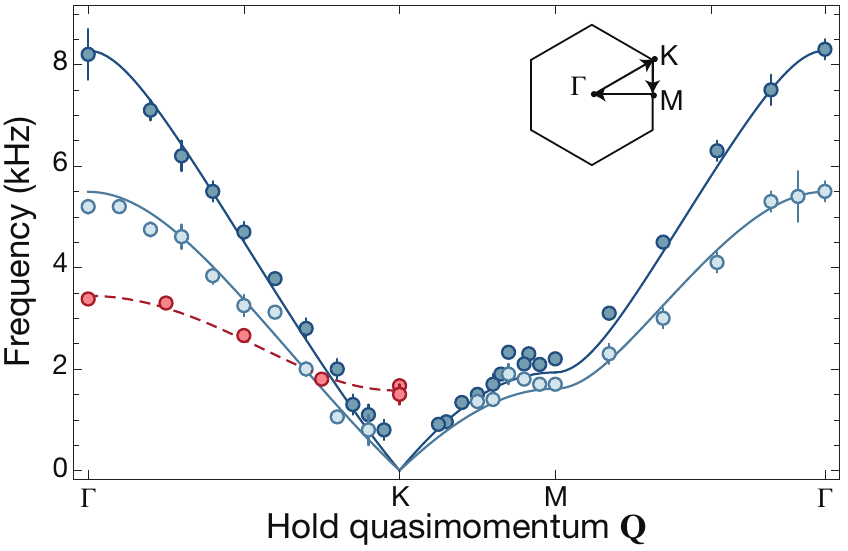}
    \caption{\textbf{Mapping the dispersion relation over the BZ.} The dispersion relation along the high-symmetry paths for lattices with: $\Delta/J$=0 and depth $V_0=$0.8$E_r$ (dark blue) or $V_0=$2.5$E_r$ (light blue); $\Delta/J$=3.1 and depth $V_0=$5.2$E_r$ (red). Theory lines show a full band structure calculation (solid) and a best-fit tight-binding model (dashed). Error bars indicate fit errors. }
	\label{Fig:S6}
\end{figure}

In addition to probing the band geometry, the interferometric sequence simultaneously reveals the dispersion relation through the frequency of the oscillation. By varying the reference quasimomentum $\vec{Q}$, we obtain the energy difference between the lower and upper bands over the entire BZ~\cite{S:Zenesini2010,S:Kling2010}. The measured dispersion along the path $\Gamma$-K-M-$\Gamma$ is shown in Fig.~\ref{Fig:S6}. This method is a convenient tool for calibrating the lattice depth and quantifying the AB-site offset. 

\vspace{2cm}
\begingroup
\renewcommand{\addcontentsline}[3]{}% Remove functionality of \addcontentsline
\renewcommand{\section}[2]{}%

%%%%%%Bibliography%%%%%%%%%
%\bibliographystyle{apsrev4-1}%apsrev4-1
%\bibliography{wilsonline_refs}

\input{wilsonline_supplementary.bbl}

\end{document}

%% file: wilsonline_main_only.bbl
%merlin.mbs apsrev4-1.bst 2010-07-25 4.21a (PWD, AO, DPC) hacked

%% file: wilsonline_supplementary.bbl
%merlin.mbs apsrev4-1.bst 2010-07-25 4.21a (PWD, AO, DPC) hacked
%Control: key (0)
%Control: author (0) dotless jnrlst
%Control: editor formatted (1) identically to author
%Control: production of article title (0) allowed
%Control: page (1) range
%Control: year (0) verbatim
%Control: production of eprint (0) enabled
%

%% file: arxiv_final.bbl
\begin{thebibliography}{46}%
\makeatletter
\providecommand \@ifxundefined [1]{%
 \@ifx{#1\undefined}
}%
\providecommand \@ifnum [1]{%
 \ifnum #1\expandafter \@firstoftwo
 \else \expandafter \@secondoftwo
 \fi
}%
\providecommand \@ifx [1]{%
 \ifx #1\expandafter \@firstoftwo
 \else \expandafter \@secondoftwo
 \fi
}%
\providecommand \natexlab [1]{#1}%
\providecommand \enquote  [1]{``#1''}%
\providecommand \bibnamefont  [1]{#1}%
\providecommand \bibfnamefont [1]{#1}%
\providecommand \citenamefont [1]{#1}%
\providecommand \href@noop [0]{\@secondoftwo}%
\providecommand \href [0]{\begingroup \@sanitize@url \@href}%
\providecommand \@href[1]{\@@startlink{#1}\@@href}%
\providecommand \@@href[1]{\endgroup#1\@@endlink}%
\providecommand \@sanitize@url [0]{\catcode `\\12\catcode `\$12\catcode
  `\&12\catcode `\#12\catcode `\^12\catcode `\_12\catcode `\%12\relax}%
\providecommand \@@startlink[1]{}%
\providecommand \@@endlink[0]{}%
\providecommand \url  [0]{\begingroup\@sanitize@url \@url }%
\providecommand \@url [1]{\endgroup\@href {#1}{\urlprefix }}%
\providecommand \urlprefix  [0]{URL }%
\providecommand \Eprint [0]{\href }%
\providecommand \doibase [0]{http://dx.doi.org/}%
\providecommand \selectlanguage [0]{\@gobble}%
\providecommand \bibinfo  [0]{\@secondoftwo}%
\providecommand \bibfield  [0]{\@secondoftwo}%
\providecommand \translation [1]{[#1]}%
\providecommand \BibitemOpen [0]{}%
\providecommand \bibitemStop [0]{}%
\providecommand \bibitemNoStop [0]{.\EOS\space}%
\providecommand \EOS [0]{\spacefactor3000\relax}%
\providecommand \BibitemShut  [1]{\csname bibitem#1\endcsname}%
\let\auto@bib@innerbib\@empty
%</preamble>
\bibitem [{\citenamefont {Xiao}\ \emph {et~al.}(2010)\citenamefont {Xiao},
  \citenamefont {Chang},\ and\ \citenamefont {Niu}}]{Xiao2010}%
  \BibitemOpen
  \bibfield  {author} {\bibinfo {author} {\bibfnamefont {Di}~\bibnamefont
  {Xiao}}, \bibinfo {author} {\bibfnamefont {Ming-Che}\ \bibnamefont {Chang}},
  \ and\ \bibinfo {author} {\bibfnamefont {Qian}\ \bibnamefont {Niu}},\
  }\bibfield  {title} {\enquote {\bibinfo {title} {{Berry phase effects on
  electronic properties}},}\ }\href {\doibase 10.1103/RevModPhys.82.1959}
  {\bibfield  {journal} {\bibinfo  {journal} {Rev. Mod. Phys.}\ }\textbf
  {\bibinfo {volume} {82}},\ \bibinfo {pages} {1959--2007} (\bibinfo {year}
  {2010})}\BibitemShut {NoStop}%
\bibitem [{\citenamefont {Thouless}\ \emph {et~al.}(1982)\citenamefont
  {Thouless}, \citenamefont {Kohmoto}, \citenamefont {Nightingale},\ and\
  \citenamefont {den Nijs}}]{Thouless1982}%
  \BibitemOpen
  \bibfield  {author} {\bibinfo {author} {\bibfnamefont {D.~J.}\ \bibnamefont
  {Thouless}}, \bibinfo {author} {\bibfnamefont {M.}~\bibnamefont {Kohmoto}},
  \bibinfo {author} {\bibfnamefont {M.~P.}\ \bibnamefont {Nightingale}}, \ and\
  \bibinfo {author} {\bibfnamefont {M.}~\bibnamefont {den Nijs}},\ }\bibfield
  {title} {\enquote {\bibinfo {title} {{Quantized Hall Conductance in a
  Two-Dimensional Periodic Potential}},}\ }\href {\doibase
  10.1103/PhysRevLett.49.405} {\bibfield  {journal} {\bibinfo  {journal} {Phys.
  Rev. Lett.}\ }\textbf {\bibinfo {volume} {49}},\ \bibinfo {pages} {405--408}
  (\bibinfo {year} {1982})}\BibitemShut {NoStop}%
\bibitem [{\citenamefont {Klitzing}\ \emph {et~al.}(1980)\citenamefont
  {Klitzing}, \citenamefont {Dorda},\ and\ \citenamefont
  {Pepper}}]{Klitzing1980}%
  \BibitemOpen
  \bibfield  {author} {\bibinfo {author} {\bibfnamefont {K.~v.}\ \bibnamefont
  {Klitzing}}, \bibinfo {author} {\bibfnamefont {G.}~\bibnamefont {Dorda}}, \
  and\ \bibinfo {author} {\bibfnamefont {M.}~\bibnamefont {Pepper}},\
  }\bibfield  {title} {\enquote {\bibinfo {title} {{New Method for
  High-Accuracy Determination of the Fine-Structure Constant Based on Quantized
  Hall Resistance}},}\ }\href {\doibase 10.1103/PhysRevLett.45.494} {\bibfield
  {journal} {\bibinfo  {journal} {Phys. Rev. Lett.}\ }\textbf {\bibinfo
  {volume} {45}},\ \bibinfo {pages} {494--497} (\bibinfo {year}
  {1980})}\BibitemShut {NoStop}%
\bibitem [{\citenamefont {Hasan}\ and\ \citenamefont {Kane}(2010)}]{Hasan2010}%
  \BibitemOpen
  \bibfield  {author} {\bibinfo {author} {\bibfnamefont {M.~Z.}\ \bibnamefont
  {Hasan}}\ and\ \bibinfo {author} {\bibfnamefont {C.~L.}\ \bibnamefont
  {Kane}},\ }\bibfield  {title} {\enquote {\bibinfo {title}
  {{\textit{Colloquium} : Topological insulators}},}\ }\href {\doibase
  10.1103/RevModPhys.82.3045} {\bibfield  {journal} {\bibinfo  {journal} {Rev.
  Mod. Phys.}\ }\textbf {\bibinfo {volume} {82}},\ \bibinfo {pages}
  {3045--3067} (\bibinfo {year} {2010})}\BibitemShut {NoStop}%
\bibitem [{\citenamefont {Qi}\ and\ \citenamefont {Zhang}(2011)}]{Qi2011}%
  \BibitemOpen
  \bibfield  {author} {\bibinfo {author} {\bibfnamefont {Xiao-Liang}\
  \bibnamefont {Qi}}\ and\ \bibinfo {author} {\bibfnamefont {Shou-Cheng}\
  \bibnamefont {Zhang}},\ }\bibfield  {title} {\enquote {\bibinfo {title}
  {{Topological insulators and superconductors}},}\ }\href {\doibase
  10.1103/RevModPhys.83.1057} {\bibfield  {journal} {\bibinfo  {journal} {Rev.
  Mod. Phys.}\ }\textbf {\bibinfo {volume} {83}},\ \bibinfo {pages}
  {1057--1110} (\bibinfo {year} {2011})}\BibitemShut {NoStop}%
\bibitem [{\citenamefont {Castro~Neto}\ \emph {et~al.}(2009)\citenamefont
  {Castro~Neto}, \citenamefont {Guinea}, \citenamefont {Peres}, \citenamefont
  {Novoselov},\ and\ \citenamefont {Geim}}]{Neto2009}%
  \BibitemOpen
  \bibfield  {author} {\bibinfo {author} {\bibfnamefont {A.~H.}\ \bibnamefont
  {Castro~Neto}}, \bibinfo {author} {\bibfnamefont {F.}~\bibnamefont {Guinea}},
  \bibinfo {author} {\bibfnamefont {N.~M.~R.}\ \bibnamefont {Peres}}, \bibinfo
  {author} {\bibfnamefont {K.~S.}\ \bibnamefont {Novoselov}}, \ and\ \bibinfo
  {author} {\bibfnamefont {A.~K.}\ \bibnamefont {Geim}},\ }\bibfield  {title}
  {\enquote {\bibinfo {title} {The electronic properties of graphene},}\ }\href
  {\doibase 10.1103/RevModPhys.81.109} {\bibfield  {journal} {\bibinfo
  {journal} {Rev. Mod. Phys.}\ }\textbf {\bibinfo {volume} {81}},\ \bibinfo
  {pages} {109--162} (\bibinfo {year} {2009})}\BibitemShut {NoStop}%
\bibitem [{\citenamefont {Yu}\ \emph {et~al.}(2011)\citenamefont {Yu},
  \citenamefont {Qi}, \citenamefont {Bernevig}, \citenamefont {Fang},\ and\
  \citenamefont {Dai}}]{Yu2011}%
  \BibitemOpen
  \bibfield  {author} {\bibinfo {author} {\bibfnamefont {Rui}\ \bibnamefont
  {Yu}}, \bibinfo {author} {\bibfnamefont {Xiao~Liang}\ \bibnamefont {Qi}},
  \bibinfo {author} {\bibfnamefont {Andrei}\ \bibnamefont {Bernevig}}, \bibinfo
  {author} {\bibfnamefont {Zhong}\ \bibnamefont {Fang}}, \ and\ \bibinfo
  {author} {\bibfnamefont {Xi}~\bibnamefont {Dai}},\ }\bibfield  {title}
  {\enquote {\bibinfo {title} {{Equivalent expression of ${\mathbb{Z}}_{2}$
  topological invariant for band insulators using the non-Abelian Berry
  connection}},}\ }\href {\doibase 10.1103/PhysRevB.84.075119} {\bibfield
  {journal} {\bibinfo  {journal} {Phys. Rev. B}\ }\textbf {\bibinfo {volume}
  {84}},\ \bibinfo {pages} {075119} (\bibinfo {year} {2011})}\BibitemShut
  {NoStop}%
\bibitem [{\citenamefont {{Alexandradinata}}\ and\ \citenamefont
  {{Bernevig}}(2014)}]{Alexandradinata2014b}%
  \BibitemOpen
  \bibfield  {author} {\bibinfo {author} {\bibfnamefont {A.}~\bibnamefont
  {{Alexandradinata}}}\ and\ \bibinfo {author} {\bibfnamefont {B.~A.}\
  \bibnamefont {{Bernevig}}},\ }\bibfield  {title} {\enquote {\bibinfo {title}
  {{Berry-phase description of Topological Crystalline Insulators}},}\
  }\href@noop {} {\bibfield  {journal} {\bibinfo  {journal} {arXiv:1409.3236}\
  } (\bibinfo {year} {2014})},\ \Eprint {http://arxiv.org/abs/1409.3236}
  {arXiv:1409.3236 [cond-mat.other]} \BibitemShut {NoStop}%
\bibitem [{\citenamefont {Alexandradinata}\ \emph {et~al.}(2014)\citenamefont
  {Alexandradinata}, \citenamefont {Dai},\ and\ \citenamefont
  {Bernevig}}]{Alexandradinata2014}%
  \BibitemOpen
  \bibfield  {author} {\bibinfo {author} {\bibfnamefont {A}~\bibnamefont
  {Alexandradinata}}, \bibinfo {author} {\bibfnamefont {Xi}~\bibnamefont
  {Dai}}, \ and\ \bibinfo {author} {\bibfnamefont {BA}~\bibnamefont
  {Bernevig}},\ }\bibfield  {title} {\enquote {\bibinfo {title} {{Wilson-loop
  characterization of inversion-symmetric topological insulators}},}\ }\href
  {http://journals.aps.org/prb/abstract/10.1103/PhysRevB.89.155114} {\bibfield
  {journal} {\bibinfo  {journal} {Phys. Rev. B}\ }\textbf {\bibinfo {volume}
  {89}},\ \bibinfo {pages} {155114} (\bibinfo {year} {2014})}\BibitemShut
  {NoStop}%
\bibitem [{\citenamefont {Grusdt}\ \emph {et~al.}(2014)\citenamefont {Grusdt},
  \citenamefont {Abanin},\ and\ \citenamefont {Demler}}]{Grusdt2014}%
  \BibitemOpen
  \bibfield  {author} {\bibinfo {author} {\bibfnamefont {F.}~\bibnamefont
  {Grusdt}}, \bibinfo {author} {\bibfnamefont {D.}~\bibnamefont {Abanin}}, \
  and\ \bibinfo {author} {\bibfnamefont {E.}~\bibnamefont {Demler}},\
  }\bibfield  {title} {\enquote {\bibinfo {title} {{Measuring
  ${\mathbb{Z}}_{2}$ topological invariants in optical lattices using
  interferometry}},}\ }\href {\doibase 10.1103/PhysRevA.89.043621} {\bibfield
  {journal} {\bibinfo  {journal} {Phys. Rev. A}\ }\textbf {\bibinfo {volume}
  {89}},\ \bibinfo {pages} {043621} (\bibinfo {year} {2014})}\BibitemShut
  {NoStop}%
\bibitem [{\citenamefont {Wilczek}\ and\ \citenamefont
  {Zee}(1984)}]{Wilczek1984}%
  \BibitemOpen
  \bibfield  {author} {\bibinfo {author} {\bibfnamefont {Frank}\ \bibnamefont
  {Wilczek}}\ and\ \bibinfo {author} {\bibfnamefont {A}~\bibnamefont {Zee}},\
  }\bibfield  {title} {\enquote {\bibinfo {title} {{Appearance of gauge
  structure in simple dynamical systems}},}\ }\href
  {http://journals.aps.org/prl/abstract/10.1103/PhysRevLett.52.2111} {\bibfield
   {journal} {\bibinfo  {journal} {Phys. Rev. Lett}\ }\textbf {\bibinfo
  {volume} {52}},\ \bibinfo {pages} {2111--2114} (\bibinfo {year}
  {1984})}\BibitemShut {NoStop}%
\bibitem [{\citenamefont {Jotzu}\ \emph {et~al.}(2014)\citenamefont {Jotzu},
  \citenamefont {Messer}, \citenamefont {Desbuquois}, \citenamefont {Lebrat},
  \citenamefont {Uehlinger}, \citenamefont {Greif},\ and\ \citenamefont
  {Esslinger}}]{Jotzu2014}%
  \BibitemOpen
  \bibfield  {author} {\bibinfo {author} {\bibfnamefont {Gregor}\ \bibnamefont
  {Jotzu}}, \bibinfo {author} {\bibfnamefont {Michael}\ \bibnamefont {Messer}},
  \bibinfo {author} {\bibfnamefont {R\'{e}mi}\ \bibnamefont {Desbuquois}},
  \bibinfo {author} {\bibfnamefont {Martin}\ \bibnamefont {Lebrat}}, \bibinfo
  {author} {\bibfnamefont {Thomas}\ \bibnamefont {Uehlinger}}, \bibinfo
  {author} {\bibfnamefont {Daniel}\ \bibnamefont {Greif}}, \ and\ \bibinfo
  {author} {\bibfnamefont {Tilman}\ \bibnamefont {Esslinger}},\ }\bibfield
  {title} {\enquote {\bibinfo {title} {{Experimental realization of the
  topological Haldane model with ultracold fermions}},}\ }\href {\doibase
  10.1038/nature13915} {\bibfield  {journal} {\bibinfo  {journal} {Nature}\
  }\textbf {\bibinfo {volume} {515}},\ \bibinfo {pages} {237--240} (\bibinfo
  {year} {2014})}\BibitemShut {NoStop}%
\bibitem [{\citenamefont {Aidelsburger}\ \emph {et~al.}(2014)\citenamefont
  {Aidelsburger}, \citenamefont {Lohse}, \citenamefont {Schweizer},
  \citenamefont {Atala}, \citenamefont {Barreiro}, \citenamefont
  {Nascimb\`{e}ne}, \citenamefont {Cooper}, \citenamefont {Bloch},\ and\
  \citenamefont {Goldman}}]{Aidelsburger2014}%
  \BibitemOpen
  \bibfield  {author} {\bibinfo {author} {\bibfnamefont {M.}~\bibnamefont
  {Aidelsburger}}, \bibinfo {author} {\bibfnamefont {M.}~\bibnamefont {Lohse}},
  \bibinfo {author} {\bibfnamefont {C.}~\bibnamefont {Schweizer}}, \bibinfo
  {author} {\bibfnamefont {M.}~\bibnamefont {Atala}}, \bibinfo {author}
  {\bibfnamefont {J.~T.}\ \bibnamefont {Barreiro}}, \bibinfo {author}
  {\bibfnamefont {S.}~\bibnamefont {Nascimb\`{e}ne}}, \bibinfo {author}
  {\bibfnamefont {N.~R.}\ \bibnamefont {Cooper}}, \bibinfo {author}
  {\bibfnamefont {I.}~\bibnamefont {Bloch}}, \ and\ \bibinfo {author}
  {\bibfnamefont {N.}~\bibnamefont {Goldman}},\ }\bibfield  {title} {\enquote
  {\bibinfo {title} {{Measuring the Chern number of Hofstadter bands with
  ultracold bosonic atoms}},}\ }\href {\doibase 10.1038/nphys3171} {\bibfield
  {journal} {\bibinfo  {journal} {Nature Physics}\ }\textbf {\bibinfo {volume}
  {11}},\ \bibinfo {pages} {162--166} (\bibinfo {year} {2014})}\BibitemShut
  {NoStop}%
\bibitem [{\citenamefont {Duca}\ \emph {et~al.}(2015)\citenamefont {Duca},
  \citenamefont {Li}, \citenamefont {Reitter}, \citenamefont {Bloch},
  \citenamefont {Schleier-Smith},\ and\ \citenamefont {Schneider}}]{Duca2015}%
  \BibitemOpen
  \bibfield  {author} {\bibinfo {author} {\bibfnamefont {L.}~\bibnamefont
  {Duca}}, \bibinfo {author} {\bibfnamefont {T.}~\bibnamefont {Li}}, \bibinfo
  {author} {\bibfnamefont {M.}~\bibnamefont {Reitter}}, \bibinfo {author}
  {\bibfnamefont {I.}~\bibnamefont {Bloch}}, \bibinfo {author} {\bibfnamefont
  {M.}~\bibnamefont {Schleier-Smith}}, \ and\ \bibinfo {author} {\bibfnamefont
  {U.}~\bibnamefont {Schneider}},\ }\bibfield  {title} {\enquote {\bibinfo
  {title} {{An Aharonov-Bohm interferometer for determining Bloch band
  topology}},}\ }\href {\doibase 10.1126/science.1259052} {\bibfield  {journal}
  {\bibinfo  {journal} {Science}\ }\textbf {\bibinfo {volume} {347}},\ \bibinfo
  {pages} {288--292} (\bibinfo {year} {2015})}\BibitemShut {NoStop}%
\bibitem [{\citenamefont {Atala}\ \emph {et~al.}(2013)\citenamefont {Atala},
  \citenamefont {Aidelsburger}, \citenamefont {Barreiro}, \citenamefont
  {Abanin}, \citenamefont {Kitagawa}, \citenamefont {Demler},\ and\
  \citenamefont {Bloch}}]{Atala2013}%
  \BibitemOpen
  \bibfield  {author} {\bibinfo {author} {\bibfnamefont {Marcos}\ \bibnamefont
  {Atala}}, \bibinfo {author} {\bibfnamefont {Monika}\ \bibnamefont
  {Aidelsburger}}, \bibinfo {author} {\bibfnamefont {Julio~T.}\ \bibnamefont
  {Barreiro}}, \bibinfo {author} {\bibfnamefont {Dmitry}\ \bibnamefont
  {Abanin}}, \bibinfo {author} {\bibfnamefont {Takuya}\ \bibnamefont
  {Kitagawa}}, \bibinfo {author} {\bibfnamefont {Eugene}\ \bibnamefont
  {Demler}}, \ and\ \bibinfo {author} {\bibfnamefont {Immanuel}\ \bibnamefont
  {Bloch}},\ }\bibfield  {title} {\enquote {\bibinfo {title} {{Direct
  measurement of the Zak phase in topological Bloch bands}},}\ }\href {\doibase
  10.1038/nphys2790} {\bibfield  {journal} {\bibinfo  {journal} {Nature
  Physics}\ }\textbf {\bibinfo {volume} {9}},\ \bibinfo {pages} {795--800}
  (\bibinfo {year} {2013})}\BibitemShut {NoStop}%
\bibitem [{\citenamefont {Zahid~Hasan}\ \emph {et~al.}(2015)\citenamefont
  {Zahid~Hasan}, \citenamefont {Xu},\ and\ \citenamefont
  {Neupane}}]{Hasan2015}%
  \BibitemOpen
  \bibfield  {author} {\bibinfo {author} {\bibfnamefont {M.}~\bibnamefont
  {Zahid~Hasan}}, \bibinfo {author} {\bibfnamefont {Su-Yang}\ \bibnamefont
  {Xu}}, \ and\ \bibinfo {author} {\bibfnamefont {Madhab}\ \bibnamefont
  {Neupane}},\ }\enquote {\bibinfo {title} {Topological insulators, topological
  dirac semimetals, topological crystalline insulators, and topological kondo
  insulators},}\ in\ \href {\doibase 10.1002/9783527681594.ch4} {\emph
  {\bibinfo {booktitle} {Topological Insulators}}}\ (\bibinfo  {publisher}
  {Wiley-VCH Verlag GmbH \& Co. KGaA},\ \bibinfo {year} {2015})\ pp.\ \bibinfo
  {pages} {55--100}\BibitemShut {NoStop}%
\bibitem [{\citenamefont {Greiner}\ \emph {et~al.}(2001)\citenamefont
  {Greiner}, \citenamefont {Bloch}, \citenamefont {Mandel}, \citenamefont
  {H\"ansch},\ and\ \citenamefont {Esslinger}}]{Greiner2001}%
  \BibitemOpen
  \bibfield  {author} {\bibinfo {author} {\bibfnamefont {Markus}\ \bibnamefont
  {Greiner}}, \bibinfo {author} {\bibfnamefont {Immanuel}\ \bibnamefont
  {Bloch}}, \bibinfo {author} {\bibfnamefont {Olaf}\ \bibnamefont {Mandel}},
  \bibinfo {author} {\bibfnamefont {Theodor~W.}\ \bibnamefont {H\"ansch}}, \
  and\ \bibinfo {author} {\bibfnamefont {Tilman}\ \bibnamefont {Esslinger}},\
  }\bibfield  {title} {\enquote {\bibinfo {title} {{Exploring Phase Coherence
  in a 2D Lattice of Bose-Einstein Condensates}},}\ }\href {\doibase
  10.1103/PhysRevLett.87.160405} {\bibfield  {journal} {\bibinfo  {journal}
  {Phys. Rev. Lett.}\ }\textbf {\bibinfo {volume} {87}},\ \bibinfo {pages}
  {160405} (\bibinfo {year} {2001})}\BibitemShut {NoStop}%
\bibitem [{\citenamefont {Ben~Dahan}\ \emph {et~al.}(1996)\citenamefont
  {Ben~Dahan}, \citenamefont {Peik}, \citenamefont {Reichel}, \citenamefont
  {Castin},\ and\ \citenamefont {Salomon}}]{Dahan1996}%
  \BibitemOpen
  \bibfield  {author} {\bibinfo {author} {\bibfnamefont {Maxime}\ \bibnamefont
  {Ben~Dahan}}, \bibinfo {author} {\bibfnamefont {Ekkehard}\ \bibnamefont
  {Peik}}, \bibinfo {author} {\bibfnamefont {Jakob}\ \bibnamefont {Reichel}},
  \bibinfo {author} {\bibfnamefont {Yvan}\ \bibnamefont {Castin}}, \ and\
  \bibinfo {author} {\bibfnamefont {Christophe}\ \bibnamefont {Salomon}},\
  }\bibfield  {title} {\enquote {\bibinfo {title} {{Bloch Oscillations of Atoms
  in an Optical Potential}},}\ }\href {\doibase 10.1103/PhysRevLett.76.4508}
  {\bibfield  {journal} {\bibinfo  {journal} {Phys. Rev. Lett.}\ }\textbf
  {\bibinfo {volume} {76}},\ \bibinfo {pages} {4508--4511} (\bibinfo {year}
  {1996})}\BibitemShut {NoStop}%
\bibitem [{SOM()}]{SOM}%
  \BibitemOpen
  \href@noop {} {}\bibinfo {note} {See Supplementary Materials on {\it Science}
  Online}\BibitemShut {NoStop}%
\bibitem [{\citenamefont {Zwanziger}\ \emph {et~al.}(1990)\citenamefont
  {Zwanziger}, \citenamefont {Koenig},\ and\ \citenamefont
  {Pines}}]{Zwanziger1990}%
  \BibitemOpen
  \bibfield  {author} {\bibinfo {author} {\bibfnamefont {J.~W.}\ \bibnamefont
  {Zwanziger}}, \bibinfo {author} {\bibfnamefont {M.}~\bibnamefont {Koenig}}, \
  and\ \bibinfo {author} {\bibfnamefont {A.}~\bibnamefont {Pines}},\ }\bibfield
   {title} {\enquote {\bibinfo {title} {{Berry's Phase}},}\ }\href {\doibase
  10.1146/annurev.pc.41.100190.003125} {\bibfield  {journal} {\bibinfo
  {journal} {Annu. Rev. Phys. Chem.}\ }\textbf {\bibinfo {volume} {41}},\
  \bibinfo {pages} {601--646} (\bibinfo {year} {1990})}\BibitemShut {NoStop}%
\bibitem [{\citenamefont {King-Smith}\ and\ \citenamefont
  {Vanderbilt}(1993)}]{Kingsmith1993}%
  \BibitemOpen
  \bibfield  {author} {\bibinfo {author} {\bibfnamefont {R.~D.}\ \bibnamefont
  {King-Smith}}\ and\ \bibinfo {author} {\bibfnamefont {David}\ \bibnamefont
  {Vanderbilt}},\ }\bibfield  {title} {\enquote {\bibinfo {title} {Theory of
  polarization of crystalline solids},}\ }\href {\doibase
  10.1103/PhysRevB.47.1651} {\bibfield  {journal} {\bibinfo  {journal} {Phys.
  Rev. B}\ }\textbf {\bibinfo {volume} {47}},\ \bibinfo {pages} {1651--1654}
  (\bibinfo {year} {1993})}\BibitemShut {NoStop}%
\bibitem [{\citenamefont {Shevchenko}\ \emph {et~al.}(2010)\citenamefont
  {Shevchenko}, \citenamefont {Ashhab},\ and\ \citenamefont
  {Nori}}]{Shevchenko2010}%
  \BibitemOpen
  \bibfield  {author} {\bibinfo {author} {\bibfnamefont {S.N.}\ \bibnamefont
  {Shevchenko}}, \bibinfo {author} {\bibfnamefont {S.}~\bibnamefont {Ashhab}},
  \ and\ \bibinfo {author} {\bibfnamefont {Franco}\ \bibnamefont {Nori}},\
  }\bibfield  {title} {\enquote {\bibinfo {title}
  {{Landau-Zener-St\"{u}ckelberg interferometry}},}\ }\href {\doibase
  10.1016/j.physrep.2010.03.002} {\bibfield  {journal} {\bibinfo  {journal}
  {Phys. Rep.}\ }\textbf {\bibinfo {volume} {492}},\ \bibinfo {pages} {1--30}
  (\bibinfo {year} {2010})}\BibitemShut {NoStop}%
\bibitem [{\citenamefont {Tarruell}\ \emph {et~al.}(2012)\citenamefont
  {Tarruell}, \citenamefont {Greif}, \citenamefont {Uehlinger}, \citenamefont
  {Jotzu},\ and\ \citenamefont {Esslinger}}]{Tarruell2012}%
  \BibitemOpen
  \bibfield  {author} {\bibinfo {author} {\bibfnamefont {Leticia}\ \bibnamefont
  {Tarruell}}, \bibinfo {author} {\bibfnamefont {Daniel}\ \bibnamefont
  {Greif}}, \bibinfo {author} {\bibfnamefont {Thomas}\ \bibnamefont
  {Uehlinger}}, \bibinfo {author} {\bibfnamefont {Gregor}\ \bibnamefont
  {Jotzu}}, \ and\ \bibinfo {author} {\bibfnamefont {Tilman}\ \bibnamefont
  {Esslinger}},\ }\bibfield  {title} {\enquote {\bibinfo {title} {{Creating,
  moving and merging Dirac points with a Fermi gas in a tunable honeycomb
  lattice.}}}\ }\href {\doibase 10.1038/nature10871} {\bibfield  {journal}
  {\bibinfo  {journal} {Nature}\ }\textbf {\bibinfo {volume} {483}},\ \bibinfo
  {pages} {302--5} (\bibinfo {year} {2012})}\BibitemShut {NoStop}%
\bibitem [{\citenamefont {Hauke}\ \emph {et~al.}(2014)\citenamefont {Hauke},
  \citenamefont {Lewenstein},\ and\ \citenamefont {Eckardt}}]{Hauke2014}%
  \BibitemOpen
  \bibfield  {author} {\bibinfo {author} {\bibfnamefont {Philipp}\ \bibnamefont
  {Hauke}}, \bibinfo {author} {\bibfnamefont {Maciej}\ \bibnamefont
  {Lewenstein}}, \ and\ \bibinfo {author} {\bibfnamefont {Andr\'e}\
  \bibnamefont {Eckardt}},\ }\bibfield  {title} {\enquote {\bibinfo {title}
  {Tomography of band insulators from quench dynamics},}\ }\href {\doibase
  10.1103/PhysRevLett.113.045303} {\bibfield  {journal} {\bibinfo  {journal}
  {Phys. Rev. Lett.}\ }\textbf {\bibinfo {volume} {113}},\ \bibinfo {pages}
  {045303} (\bibinfo {year} {2014})}\BibitemShut {NoStop}%
\bibitem [{\citenamefont {Alba}\ \emph {et~al.}(2011)\citenamefont {Alba},
  \citenamefont {Fernandez-Gonzalvo}, \citenamefont {Mur-Petit}, \citenamefont
  {Pachos},\ and\ \citenamefont {Garcia-Ripoll}}]{Alba2011}%
  \BibitemOpen
  \bibfield  {author} {\bibinfo {author} {\bibfnamefont {E.}~\bibnamefont
  {Alba}}, \bibinfo {author} {\bibfnamefont {X.}~\bibnamefont
  {Fernandez-Gonzalvo}}, \bibinfo {author} {\bibfnamefont {J.}~\bibnamefont
  {Mur-Petit}}, \bibinfo {author} {\bibfnamefont {J.~K.}\ \bibnamefont
  {Pachos}}, \ and\ \bibinfo {author} {\bibfnamefont {J.~J.}\ \bibnamefont
  {Garcia-Ripoll}},\ }\bibfield  {title} {\enquote {\bibinfo {title} {Seeing
  topological order in time-of-flight measurements},}\ }\href {\doibase
  10.1103/PhysRevLett.107.235301} {\bibfield  {journal} {\bibinfo  {journal}
  {Phys. Rev. Lett.}\ }\textbf {\bibinfo {volume} {107}},\ \bibinfo {pages}
  {235301} (\bibinfo {year} {2011})}\BibitemShut {NoStop}%
\bibitem [{\citenamefont {Zenesini}\ \emph {et~al.}(2010)\citenamefont
  {Zenesini}, \citenamefont {Ciampini}, \citenamefont {Morsch},\ and\
  \citenamefont {Arimondo}}]{Zenesini2010}%
  \BibitemOpen
  \bibfield  {author} {\bibinfo {author} {\bibfnamefont {A.}~\bibnamefont
  {Zenesini}}, \bibinfo {author} {\bibfnamefont {D.}~\bibnamefont {Ciampini}},
  \bibinfo {author} {\bibfnamefont {O.}~\bibnamefont {Morsch}}, \ and\ \bibinfo
  {author} {\bibfnamefont {E.}~\bibnamefont {Arimondo}},\ }\bibfield  {title}
  {\enquote {\bibinfo {title} {{Observation of St\"uckelberg oscillations in
  accelerated optical lattices}},}\ }\href {\doibase
  10.1103/PhysRevA.82.065601} {\bibfield  {journal} {\bibinfo  {journal} {Phys.
  Rev. A}\ }\textbf {\bibinfo {volume} {82}},\ \bibinfo {pages} {065601}
  (\bibinfo {year} {2010})}\BibitemShut {NoStop}%
\bibitem [{\citenamefont {Kling}\ \emph {et~al.}(2010)\citenamefont {Kling},
  \citenamefont {Salger}, \citenamefont {Grossert},\ and\ \citenamefont
  {Weitz}}]{Kling2010}%
  \BibitemOpen
  \bibfield  {author} {\bibinfo {author} {\bibfnamefont {Sebastian}\
  \bibnamefont {Kling}}, \bibinfo {author} {\bibfnamefont {Tobias}\
  \bibnamefont {Salger}}, \bibinfo {author} {\bibfnamefont {Christopher}\
  \bibnamefont {Grossert}}, \ and\ \bibinfo {author} {\bibfnamefont {Martin}\
  \bibnamefont {Weitz}},\ }\bibfield  {title} {\enquote {\bibinfo {title}
  {{Atomic Bloch-Zener Oscillations and St\"{u}ckelberg Interferometry in
  Optical Lattices}},}\ }\href {\doibase 10.1103/PhysRevLett.105.215301}
  {\bibfield  {journal} {\bibinfo  {journal} {Phys. Rev. Lett.}\ }\textbf
  {\bibinfo {volume} {105}},\ \bibinfo {pages} {215301} (\bibinfo {year}
  {2010})}\BibitemShut {NoStop}%
\bibitem [{\citenamefont {Lim}\ \emph {et~al.}(2014)\citenamefont {Lim},
  \citenamefont {Fuchs},\ and\ \citenamefont {Montambaux}}]{Lim2014}%
  \BibitemOpen
  \bibfield  {author} {\bibinfo {author} {\bibfnamefont {Lih-King}\
  \bibnamefont {Lim}}, \bibinfo {author} {\bibfnamefont {Jean-No\"el}\
  \bibnamefont {Fuchs}}, \ and\ \bibinfo {author} {\bibfnamefont {Gilles}\
  \bibnamefont {Montambaux}},\ }\bibfield  {title} {\enquote {\bibinfo {title}
  {Mass and chirality inversion of a dirac cone pair in st\"uckelberg
  interferometry},}\ }\href {\doibase 10.1103/PhysRevLett.112.155302}
  {\bibfield  {journal} {\bibinfo  {journal} {Phys. Rev. Lett.}\ }\textbf
  {\bibinfo {volume} {112}},\ \bibinfo {pages} {155302} (\bibinfo {year}
  {2014})}\BibitemShut {NoStop}%
\bibitem [{\citenamefont {Lim}\ \emph {et~al.}(2015)\citenamefont {Lim},
  \citenamefont {Fuchs},\ and\ \citenamefont {Montambaux}}]{Lim2015}%
  \BibitemOpen
  \bibfield  {author} {\bibinfo {author} {\bibfnamefont {Lih-King}\
  \bibnamefont {Lim}}, \bibinfo {author} {\bibfnamefont {Jean-No\"el}\
  \bibnamefont {Fuchs}}, \ and\ \bibinfo {author} {\bibfnamefont {Gilles}\
  \bibnamefont {Montambaux}},\ }\bibfield  {title} {\enquote {\bibinfo {title}
  {Geometric phase in st\"uckelberg interferometry},}\ }\href {\doibase
  10.1103/PhysRevA.91.042119} {\bibfield  {journal} {\bibinfo  {journal} {Phys.
  Rev. A}\ }\textbf {\bibinfo {volume} {91}},\ \bibinfo {pages} {042119}
  (\bibinfo {year} {2015})}\BibitemShut {NoStop}%
\bibitem [{\citenamefont {Baur}\ \emph {et~al.}(2014)\citenamefont {Baur},
  \citenamefont {Schleier-Smith},\ and\ \citenamefont {Cooper}}]{Baur2014}%
  \BibitemOpen
  \bibfield  {author} {\bibinfo {author} {\bibfnamefont {Stefan~K.}\
  \bibnamefont {Baur}}, \bibinfo {author} {\bibfnamefont {Monika~H.}\
  \bibnamefont {Schleier-Smith}}, \ and\ \bibinfo {author} {\bibfnamefont
  {Nigel~R.}\ \bibnamefont {Cooper}},\ }\bibfield  {title} {\enquote {\bibinfo
  {title} {Dynamic optical superlattices with topological bands},}\ }\href
  {\doibase 10.1103/PhysRevA.89.051605} {\bibfield  {journal} {\bibinfo
  {journal} {Phys. Rev. A}\ }\textbf {\bibinfo {volume} {89}},\ \bibinfo
  {pages} {051605} (\bibinfo {year} {2014})}\BibitemShut {NoStop}%
\bibitem [{\citenamefont {Abanin}\ \emph {et~al.}(2013)\citenamefont {Abanin},
  \citenamefont {Kitagawa}, \citenamefont {Bloch},\ and\ \citenamefont
  {Demler}}]{Abanin2013}%
  \BibitemOpen
  \bibfield  {author} {\bibinfo {author} {\bibfnamefont {Dmitry~A.}\
  \bibnamefont {Abanin}}, \bibinfo {author} {\bibfnamefont {Takuya}\
  \bibnamefont {Kitagawa}}, \bibinfo {author} {\bibfnamefont {Immanuel}\
  \bibnamefont {Bloch}}, \ and\ \bibinfo {author} {\bibfnamefont {Eugene}\
  \bibnamefont {Demler}},\ }\bibfield  {title} {\enquote {\bibinfo {title}
  {{Interferometric Approach to Measuring Band Topology in 2D Optical
  Lattices}},}\ }\href {\doibase 10.1103/PhysRevLett.110.165304} {\bibfield
  {journal} {\bibinfo  {journal} {Phys. Rev. Lett.}\ }\textbf {\bibinfo
  {volume} {110}},\ \bibinfo {pages} {165304} (\bibinfo {year}
  {2013})}\BibitemShut {NoStop}%
\bibitem [{\citenamefont {Poyatos}\ \emph {et~al.}(1997)\citenamefont
  {Poyatos}, \citenamefont {Cirac},\ and\ \citenamefont
  {Zoller}}]{Poyatos1997}%
  \BibitemOpen
  \bibfield  {author} {\bibinfo {author} {\bibfnamefont {J.~F.}\ \bibnamefont
  {Poyatos}}, \bibinfo {author} {\bibfnamefont {J.~I.}\ \bibnamefont {Cirac}},
  \ and\ \bibinfo {author} {\bibfnamefont {P.}~\bibnamefont {Zoller}},\
  }\bibfield  {title} {\enquote {\bibinfo {title} {Complete characterization of
  a quantum process: The two-bit quantum gate},}\ }\href {\doibase
  10.1103/PhysRevLett.78.390} {\bibfield  {journal} {\bibinfo  {journal} {Phys.
  Rev. Lett.}\ }\textbf {\bibinfo {volume} {78}},\ \bibinfo {pages} {390--393}
  (\bibinfo {year} {1997})}\BibitemShut {NoStop}%
\bibitem [{\citenamefont {Soluyanov}\ and\ \citenamefont
  {Vanderbilt}(2011)}]{Soluyanov2011}%
  \BibitemOpen
  \bibfield  {author} {\bibinfo {author} {\bibfnamefont {Alexey~A.}\
  \bibnamefont {Soluyanov}}\ and\ \bibinfo {author} {\bibfnamefont {David}\
  \bibnamefont {Vanderbilt}},\ }\bibfield  {title} {\enquote {\bibinfo {title}
  {{Wannier representation of ${\mathbb{Z}}_{2}$ topological insulators}},}\
  }\href {\doibase 10.1103/PhysRevB.83.035108} {\bibfield  {journal} {\bibinfo
  {journal} {Phys. Rev. B}\ }\textbf {\bibinfo {volume} {83}},\ \bibinfo
  {pages} {035108} (\bibinfo {year} {2011})}\BibitemShut {NoStop}%
\bibitem [{\citenamefont {Aidelsburger}\ \emph {et~al.}(2013)\citenamefont
  {Aidelsburger}, \citenamefont {Atala}, \citenamefont {Lohse}, \citenamefont
  {Barreiro}, \citenamefont {Paredes},\ and\ \citenamefont
  {Bloch}}]{Aidelsburger2013}%
  \BibitemOpen
  \bibfield  {author} {\bibinfo {author} {\bibfnamefont {M.}~\bibnamefont
  {Aidelsburger}}, \bibinfo {author} {\bibfnamefont {M.}~\bibnamefont {Atala}},
  \bibinfo {author} {\bibfnamefont {M.}~\bibnamefont {Lohse}}, \bibinfo
  {author} {\bibfnamefont {J.~T.}\ \bibnamefont {Barreiro}}, \bibinfo {author}
  {\bibfnamefont {B.}~\bibnamefont {Paredes}}, \ and\ \bibinfo {author}
  {\bibfnamefont {I.}~\bibnamefont {Bloch}},\ }\bibfield  {title} {\enquote
  {\bibinfo {title} {{Realization of the Hofstadter Hamiltonian with Ultracold
  Atoms in Optical Lattices}},}\ }\href {\doibase
  10.1103/PhysRevLett.111.185301} {\bibfield  {journal} {\bibinfo  {journal}
  {Phys. Rev. Lett.}\ }\textbf {\bibinfo {volume} {111}},\ \bibinfo {pages}
  {185301} (\bibinfo {year} {2013})}\BibitemShut {NoStop}%
\bibitem [{\citenamefont {Miyake}\ \emph {et~al.}(2013)\citenamefont {Miyake},
  \citenamefont {Siviloglou}, \citenamefont {Kennedy}, \citenamefont {Burton},\
  and\ \citenamefont {Ketterle}}]{Miyake2013}%
  \BibitemOpen
  \bibfield  {author} {\bibinfo {author} {\bibfnamefont {Hirokazu}\
  \bibnamefont {Miyake}}, \bibinfo {author} {\bibfnamefont {Georgios~A.}\
  \bibnamefont {Siviloglou}}, \bibinfo {author} {\bibfnamefont {Colin~J.}\
  \bibnamefont {Kennedy}}, \bibinfo {author} {\bibfnamefont {William~Cody}\
  \bibnamefont {Burton}}, \ and\ \bibinfo {author} {\bibfnamefont {Wolfgang}\
  \bibnamefont {Ketterle}},\ }\bibfield  {title} {\enquote {\bibinfo {title}
  {{Realizing the Harper Hamiltonian with Laser-Assisted Tunneling in Optical
  Lattices}},}\ }\href {\doibase 10.1103/PhysRevLett.111.185302} {\bibfield
  {journal} {\bibinfo  {journal} {Phys. Rev. Lett.}\ }\textbf {\bibinfo
  {volume} {111}},\ \bibinfo {pages} {185302} (\bibinfo {year}
  {2013})}\BibitemShut {NoStop}%
\bibitem [{\citenamefont {{Lindner}}\ \emph {et~al.}(2011)\citenamefont
  {{Lindner}}, \citenamefont {{Refael}},\ and\ \citenamefont
  {{Galitski}}}]{Lindner2011}%
  \BibitemOpen
  \bibfield  {author} {\bibinfo {author} {\bibfnamefont {N.~H.}\ \bibnamefont
  {{Lindner}}}, \bibinfo {author} {\bibfnamefont {G.}~\bibnamefont {{Refael}}},
  \ and\ \bibinfo {author} {\bibfnamefont {V.}~\bibnamefont {{Galitski}}},\
  }\bibfield  {title} {\enquote {\bibinfo {title} {{Floquet topological
  insulator in semiconductor quantum wells}},}\ }\href {\doibase
  10.1038/nphys1926} {\bibfield  {journal} {\bibinfo  {journal} {Nature
  Physics}\ }\textbf {\bibinfo {volume} {7}},\ \bibinfo {pages} {490--495}
  (\bibinfo {year} {2011})}\BibitemShut {NoStop}%
\bibitem [{\citenamefont {Parker}\ \emph {et~al.}(2013)\citenamefont {Parker},
  \citenamefont {Ha},\ and\ \citenamefont {Chin}}]{Parker2013}%
  \BibitemOpen
  \bibfield  {author} {\bibinfo {author} {\bibfnamefont {Colin~V.}\
  \bibnamefont {Parker}}, \bibinfo {author} {\bibfnamefont {Li-Chung}\
  \bibnamefont {Ha}}, \ and\ \bibinfo {author} {\bibfnamefont {Cheng}\
  \bibnamefont {Chin}},\ }\bibfield  {title} {\enquote {\bibinfo {title}
  {Direct observation of effective ferromagnetic domains of cold atoms in a
  shaken optical lattice},}\ }\href {http://dx.doi.org/10.1038/nphys2789}
  {\bibfield  {journal} {\bibinfo  {journal} {Nature Physics}\ }\textbf
  {\bibinfo {volume} {9}},\ \bibinfo {pages} {769--774} (\bibinfo {year}
  {2013})},\ \bibinfo {note} {letter}\BibitemShut {NoStop}%
\bibitem [{\citenamefont {Lin}\ \emph {et~al.}(2009)\citenamefont {Lin},
  \citenamefont {Compton}, \citenamefont {Jimenez-Garcia}, \citenamefont
  {Porto},\ and\ \citenamefont {Spielman}}]{Lin2009}%
  \BibitemOpen
  \bibfield  {author} {\bibinfo {author} {\bibfnamefont {Y.~J.}\ \bibnamefont
  {Lin}}, \bibinfo {author} {\bibfnamefont {R.~L.}\ \bibnamefont {Compton}},
  \bibinfo {author} {\bibfnamefont {K.}~\bibnamefont {Jimenez-Garcia}},
  \bibinfo {author} {\bibfnamefont {J.~V.}\ \bibnamefont {Porto}}, \ and\
  \bibinfo {author} {\bibfnamefont {I.~B.}\ \bibnamefont {Spielman}},\
  }\bibfield  {title} {\enquote {\bibinfo {title} {{Synthetic magnetic fields
  for ultracold neutral atoms}},}\ }\href {\doibase 10.1038/nature08609}
  {\bibfield  {journal} {\bibinfo  {journal} {Nature}\ }\textbf {\bibinfo
  {volume} {462}},\ \bibinfo {pages} {628--632} (\bibinfo {year}
  {2009})}\BibitemShut {NoStop}%
\bibitem [{\citenamefont {Cooper}(2011)}]{Cooper2011}%
  \BibitemOpen
  \bibfield  {author} {\bibinfo {author} {\bibfnamefont {N.~R.}\ \bibnamefont
  {Cooper}},\ }\bibfield  {title} {\enquote {\bibinfo {title} {{Optical Flux
  Lattices for Ultracold Atomic Gases}},}\ }\href {\doibase
  10.1103/PhysRevLett.106.175301} {\bibfield  {journal} {\bibinfo  {journal}
  {Phys. Rev. Lett.}\ }\textbf {\bibinfo {volume} {106}},\ \bibinfo {pages}
  {175301} (\bibinfo {year} {2011})}\BibitemShut {NoStop}%
\bibitem [{\citenamefont {Dalibard}\ \emph {et~al.}(2011)\citenamefont
  {Dalibard}, \citenamefont {Gerbier}, \citenamefont
  {Juzeli\ifmmode~\bar{u}\else \={u}\fi{}nas},\ and\ \citenamefont
  {\"Ohberg}}]{Dalibard2011}%
  \BibitemOpen
  \bibfield  {author} {\bibinfo {author} {\bibfnamefont {Jean}\ \bibnamefont
  {Dalibard}}, \bibinfo {author} {\bibfnamefont {Fabrice}\ \bibnamefont
  {Gerbier}}, \bibinfo {author} {\bibfnamefont {Gediminas}\ \bibnamefont
  {Juzeli\ifmmode~\bar{u}\else \={u}\fi{}nas}}, \ and\ \bibinfo {author}
  {\bibfnamefont {Patrik}\ \bibnamefont {\"Ohberg}},\ }\bibfield  {title}
  {\enquote {\bibinfo {title} {{\textit{Colloquium} : Artificial gauge
  potentials for neutral atoms}},}\ }\href {\doibase
  10.1103/RevModPhys.83.1523} {\bibfield  {journal} {\bibinfo  {journal} {Rev.
  Mod. Phys.}\ }\textbf {\bibinfo {volume} {83}},\ \bibinfo {pages}
  {1523--1543} (\bibinfo {year} {2011})}\BibitemShut {NoStop}%
\bibitem [{\citenamefont {B\'eri}\ and\ \citenamefont
  {Cooper}(2011)}]{Beri2011}%
  \BibitemOpen
  \bibfield  {author} {\bibinfo {author} {\bibfnamefont {B.}~\bibnamefont
  {B\'eri}}\ and\ \bibinfo {author} {\bibfnamefont {N.~R.}\ \bibnamefont
  {Cooper}},\ }\bibfield  {title} {\enquote {\bibinfo {title}
  {${\mathbb{z}}_{2}$ topological insulators in ultracold atomic gases},}\
  }\href {\doibase 10.1103/PhysRevLett.107.145301} {\bibfield  {journal}
  {\bibinfo  {journal} {Phys. Rev. Lett.}\ }\textbf {\bibinfo {volume} {107}},\
  \bibinfo {pages} {145301} (\bibinfo {year} {2011})}\BibitemShut {NoStop}%
\bibitem [{\citenamefont {Lin}\ \emph {et~al.}(2011)\citenamefont {Lin},
  \citenamefont {Jimenez-Garcia},\ and\ \citenamefont {Spielman}}]{Lin2011}%
  \BibitemOpen
  \bibfield  {author} {\bibinfo {author} {\bibfnamefont {Y.-J.}\ \bibnamefont
  {Lin}}, \bibinfo {author} {\bibfnamefont {K.}~\bibnamefont {Jimenez-Garcia}},
  \ and\ \bibinfo {author} {\bibfnamefont {I.~B.}\ \bibnamefont {Spielman}},\
  }\bibfield  {title} {\enquote {\bibinfo {title} {Spin-orbit-coupled
  bose-einstein condensates},}\ }\href {\doibase 10.1038/nature09887}
  {\bibfield  {journal} {\bibinfo  {journal} {Nature}\ }\textbf {\bibinfo
  {volume} {471}},\ \bibinfo {pages} {83--86} (\bibinfo {year}
  {2011})}\BibitemShut {NoStop}%
\bibitem [{\citenamefont {Goldman}\ \emph {et~al.}(2010)\citenamefont
  {Goldman}, \citenamefont {Satija}, \citenamefont {Nikolic}, \citenamefont
  {Bermudez}, \citenamefont {Martin-Delgado}, \citenamefont {Lewenstein},\ and\
  \citenamefont {Spielman}}]{Goldman2010}%
  \BibitemOpen
  \bibfield  {author} {\bibinfo {author} {\bibfnamefont {N.}~\bibnamefont
  {Goldman}}, \bibinfo {author} {\bibfnamefont {I.}~\bibnamefont {Satija}},
  \bibinfo {author} {\bibfnamefont {P.}~\bibnamefont {Nikolic}}, \bibinfo
  {author} {\bibfnamefont {A.}~\bibnamefont {Bermudez}}, \bibinfo {author}
  {\bibfnamefont {M.~A.}\ \bibnamefont {Martin-Delgado}}, \bibinfo {author}
  {\bibfnamefont {M.}~\bibnamefont {Lewenstein}}, \ and\ \bibinfo {author}
  {\bibfnamefont {I.~B.}\ \bibnamefont {Spielman}},\ }\bibfield  {title}
  {\enquote {\bibinfo {title} {Realistic time-reversal invariant topological
  insulators with neutral atoms},}\ }\href {\doibase
  10.1103/PhysRevLett.105.255302} {\bibfield  {journal} {\bibinfo  {journal}
  {Phys. Rev. Lett.}\ }\textbf {\bibinfo {volume} {105}},\ \bibinfo {pages}
  {255302} (\bibinfo {year} {2010})}\BibitemShut {NoStop}%
\bibitem [{\citenamefont {Liu}\ \emph {et~al.}(2010)\citenamefont {Liu},
  \citenamefont {Zhu}, \citenamefont {Jiang}, \citenamefont {Sun},\ and\
  \citenamefont {Liu}}]{Liu2010}%
  \BibitemOpen
  \bibfield  {author} {\bibinfo {author} {\bibfnamefont {Guocai}\ \bibnamefont
  {Liu}}, \bibinfo {author} {\bibfnamefont {Shi-Liang}\ \bibnamefont {Zhu}},
  \bibinfo {author} {\bibfnamefont {Shaojian}\ \bibnamefont {Jiang}}, \bibinfo
  {author} {\bibfnamefont {Fadi}\ \bibnamefont {Sun}}, \ and\ \bibinfo {author}
  {\bibfnamefont {W.~M.}\ \bibnamefont {Liu}},\ }\bibfield  {title} {\enquote
  {\bibinfo {title} {Simulating and detecting the quantum spin hall effect in
  the kagome optical lattice},}\ }\href {\doibase 10.1103/PhysRevA.82.053605}
  {\bibfield  {journal} {\bibinfo  {journal} {Phys. Rev. A}\ }\textbf {\bibinfo
  {volume} {82}},\ \bibinfo {pages} {053605} (\bibinfo {year}
  {2010})}\BibitemShut {NoStop}%
\bibitem [{\citenamefont {Kennedy}\ \emph {et~al.}(2013)\citenamefont
  {Kennedy}, \citenamefont {Siviloglou}, \citenamefont {Miyake}, \citenamefont
  {Burton},\ and\ \citenamefont {Ketterle}}]{Kennedy2013}%
  \BibitemOpen
  \bibfield  {author} {\bibinfo {author} {\bibfnamefont {Colin~J.}\
  \bibnamefont {Kennedy}}, \bibinfo {author} {\bibfnamefont {Georgios~A.}\
  \bibnamefont {Siviloglou}}, \bibinfo {author} {\bibfnamefont {Hirokazu}\
  \bibnamefont {Miyake}}, \bibinfo {author} {\bibfnamefont {William~Cody}\
  \bibnamefont {Burton}}, \ and\ \bibinfo {author} {\bibfnamefont {Wolfgang}\
  \bibnamefont {Ketterle}},\ }\bibfield  {title} {\enquote {\bibinfo {title}
  {Spin-orbit coupling and quantum spin hall effect for neutral atoms without
  spin flips},}\ }\href {\doibase 10.1103/PhysRevLett.111.225301} {\bibfield
  {journal} {\bibinfo  {journal} {Phys. Rev. Lett.}\ }\textbf {\bibinfo
  {volume} {111}},\ \bibinfo {pages} {225301} (\bibinfo {year}
  {2013})}\BibitemShut {NoStop}%
\bibitem [{\citenamefont {Mei}\ \emph {et~al.}(2012)\citenamefont {Mei},
  \citenamefont {Zhu}, \citenamefont {Zhang}, \citenamefont {Oh},\ and\
  \citenamefont {Goldman}}]{Mei2012}%
  \BibitemOpen
  \bibfield  {author} {\bibinfo {author} {\bibfnamefont {Feng}\ \bibnamefont
  {Mei}}, \bibinfo {author} {\bibfnamefont {Shi-Liang}\ \bibnamefont {Zhu}},
  \bibinfo {author} {\bibfnamefont {Zhi-Ming}\ \bibnamefont {Zhang}}, \bibinfo
  {author} {\bibfnamefont {C.~H.}\ \bibnamefont {Oh}}, \ and\ \bibinfo {author}
  {\bibfnamefont {N.}~\bibnamefont {Goldman}},\ }\bibfield  {title} {\enquote
  {\bibinfo {title} {Simulating ${Z}_{2}$ topological insulators with cold
  atoms in a one-dimensional optical lattice},}\ }\href {\doibase
  10.1103/PhysRevA.85.013638} {\bibfield  {journal} {\bibinfo  {journal} {Phys.
  Rev. A}\ }\textbf {\bibinfo {volume} {85}},\ \bibinfo {pages} {013638}
  (\bibinfo {year} {2012})}\BibitemShut {NoStop}%
\end{thebibliography}

\begin{thebibliography}{12}%
\makeatletter
\providecommand \@ifxundefined [1]{%
 \@ifx{#1\undefined}
}%
\providecommand \@ifnum [1]{%
 \ifnum #1\expandafter \@firstoftwo
 \else \expandafter \@secondoftwo
 \fi
}%
\providecommand \@ifx [1]{%
 \ifx #1\expandafter \@firstoftwo
 \else \expandafter \@secondoftwo
 \fi
}%
\providecommand \natexlab [1]{#1}%
\providecommand \enquote  [1]{``#1''}%
\providecommand \bibnamefont  [1]{#1}%
\providecommand \bibfnamefont [1]{#1}%
\providecommand \citenamefont [1]{#1}%
\providecommand \href@noop [0]{\@secondoftwo}%
\providecommand \href [0]{\begingroup \@sanitize@url \@href}%
\providecommand \@href[1]{\@@startlink{#1}\@@href}%
\providecommand \@@href[1]{\endgroup#1\@@endlink}%
\providecommand \@sanitize@url [0]{\catcode `\\12\catcode `\$12\catcode
  `\&12\catcode `\#12\catcode `\^12\catcode `\_12\catcode `\%12\relax}%
\providecommand \@@startlink[1]{}%
\providecommand \@@endlink[0]{}%
\providecommand \url  [0]{\begingroup\@sanitize@url \@url }%
\providecommand \@url [1]{\endgroup\@href {#1}{\urlprefix }}%
\providecommand \urlprefix  [0]{URL }%
\providecommand \Eprint [0]{\href }%
\providecommand \doibase [0]{http://dx.doi.org/}%
\providecommand \selectlanguage [0]{\@gobble}%
\providecommand \bibinfo  [0]{\@secondoftwo}%
\providecommand \bibfield  [0]{\@secondoftwo}%
\providecommand \translation [1]{[#1]}%
\providecommand \BibitemOpen [0]{}%
\providecommand \bibitemStop [0]{}%
\providecommand \bibitemNoStop [0]{.\EOS\space}%
\providecommand \EOS [0]{\spacefactor3000\relax}%
\providecommand \BibitemShut  [1]{\csname bibitem#1\endcsname}%
\let\auto@bib@innerbib\@empty
%</preamble>
\bibitem [{\citenamefont {Wilczek}\ and\ \citenamefont
  {Zee}(1984)}]{S:Wilczek1984}%
  \BibitemOpen
  \bibfield  {author} {\bibinfo {author} {\bibfnamefont {Frank}\ \bibnamefont
  {Wilczek}}\ and\ \bibinfo {author} {\bibfnamefont {A}~\bibnamefont {Zee}},\
  }\bibfield  {title} {\enquote {\bibinfo {title} {{Appearance of gauge
  structure in simple dynamical systems}},}\ }\href
  {http://journals.aps.org/prl/abstract/10.1103/PhysRevLett.52.2111} {\bibfield
   {journal} {\bibinfo  {journal} {Phys. Rev. Lett}\ }\textbf {\bibinfo
  {volume} {52}},\ \bibinfo {pages} {2111--2114} (\bibinfo {year}
  {1984})}\BibitemShut {NoStop}%
\bibitem [{\citenamefont {Semenoff}(1984)}]{S:Semenoff84}%
  \BibitemOpen
  \bibfield  {author} {\bibinfo {author} {\bibfnamefont {Gordon~W.}\
  \bibnamefont {Semenoff}},\ }\bibfield  {title} {\enquote {\bibinfo {title}
  {{Condensed-Matter Simulation of a Three-Dimensional Anomaly}},}\ }\href
  {\doibase 10.1103/PhysRevLett.53.2449} {\bibfield  {journal} {\bibinfo
  {journal} {Phys. Rev. Lett.}\ }\textbf {\bibinfo {volume} {53}},\ \bibinfo
  {pages} {2449--2452} (\bibinfo {year} {1984})}\BibitemShut {NoStop}%
\bibitem [{\citenamefont {Kohn}(1959)}]{S:Kohn1959}%
  \BibitemOpen
  \bibfield  {author} {\bibinfo {author} {\bibfnamefont {W.}~\bibnamefont
  {Kohn}},\ }\bibfield  {title} {\enquote {\bibinfo {title} {{Analytic
  Properties of Bloch Waves and Wannier Functions}},}\ }\href {\doibase
  10.1103/PhysRev.115.809} {\bibfield  {journal} {\bibinfo  {journal} {Phys.
  Rev.}\ }\textbf {\bibinfo {volume} {115}},\ \bibinfo {pages} {809--821}
  (\bibinfo {year} {1959})}\BibitemShut {NoStop}%
\bibitem [{\citenamefont {Yu}\ \emph {et~al.}(2011)\citenamefont {Yu},
  \citenamefont {Qi}, \citenamefont {Bernevig}, \citenamefont {Fang},\ and\
  \citenamefont {Dai}}]{S:Yu2011}%
  \BibitemOpen
  \bibfield  {author} {\bibinfo {author} {\bibfnamefont {Rui}\ \bibnamefont
  {Yu}}, \bibinfo {author} {\bibfnamefont {Xiao~Liang}\ \bibnamefont {Qi}},
  \bibinfo {author} {\bibfnamefont {Andrei}\ \bibnamefont {Bernevig}}, \bibinfo
  {author} {\bibfnamefont {Zhong}\ \bibnamefont {Fang}}, \ and\ \bibinfo
  {author} {\bibfnamefont {Xi}~\bibnamefont {Dai}},\ }\bibfield  {title}
  {\enquote {\bibinfo {title} {{Equivalent expression of ${\mathbb{Z}}_{2}$
  topological invariant for band insulators using the non-Abelian Berry
  connection}},}\ }\href {\doibase 10.1103/PhysRevB.84.075119} {\bibfield
  {journal} {\bibinfo  {journal} {Phys. Rev. B}\ }\textbf {\bibinfo {volume}
  {84}},\ \bibinfo {pages} {075119} (\bibinfo {year} {2011})}\BibitemShut
  {NoStop}%
\bibitem [{\citenamefont {Peskin}\ and\ \citenamefont
  {Schroeder}(1995)}]{S:Peskin1995}%
  \BibitemOpen
  \bibfield  {author} {\bibinfo {author} {\bibfnamefont {M.~E.}\ \bibnamefont
  {Peskin}}\ and\ \bibinfo {author} {\bibfnamefont {D.~V.}\ \bibnamefont
  {Schroeder}},\ }\href@noop {} {\emph {\bibinfo {title} {An Introduction to
  Quantum Field Theory (Frontiers in Physics)}}}\ (\bibinfo  {publisher}
  {Westview Press},\ \bibinfo {year} {1995})\BibitemShut {NoStop}%
\bibitem [{\citenamefont {Zee}(1988)}]{S:Zee1988}%
  \BibitemOpen
  \bibfield  {author} {\bibinfo {author} {\bibfnamefont {A.}~\bibnamefont
  {Zee}},\ }\bibfield  {title} {\enquote {\bibinfo {title} {Non-abelian gauge
  structure in nuclear quadrupole resonance},}\ }\href {\doibase
  10.1103/PhysRevA.38.1} {\bibfield  {journal} {\bibinfo  {journal} {Phys. Rev.
  A}\ }\textbf {\bibinfo {volume} {38}},\ \bibinfo {pages} {1--6} (\bibinfo
  {year} {1988})}\BibitemShut {NoStop}%
\bibitem [{\citenamefont {Thouless}\ \emph {et~al.}(1982)\citenamefont
  {Thouless}, \citenamefont {Kohmoto}, \citenamefont {Nightingale},\ and\
  \citenamefont {den Nijs}}]{S:Thouless1982}%
  \BibitemOpen
  \bibfield  {author} {\bibinfo {author} {\bibfnamefont {D.~J.}\ \bibnamefont
  {Thouless}}, \bibinfo {author} {\bibfnamefont {M.}~\bibnamefont {Kohmoto}},
  \bibinfo {author} {\bibfnamefont {M.~P.}\ \bibnamefont {Nightingale}}, \ and\
  \bibinfo {author} {\bibfnamefont {M.}~\bibnamefont {den Nijs}},\ }\bibfield
  {title} {\enquote {\bibinfo {title} {{Quantized Hall Conductance in a
  Two-Dimensional Periodic Potential}},}\ }\href {\doibase
  10.1103/PhysRevLett.49.405} {\bibfield  {journal} {\bibinfo  {journal} {Phys.
  Rev. Lett.}\ }\textbf {\bibinfo {volume} {49}},\ \bibinfo {pages} {405--408}
  (\bibinfo {year} {1982})}\BibitemShut {NoStop}%
\bibitem [{\citenamefont {Zak}(1989)}]{S:Zak1989}%
  \BibitemOpen
  \bibfield  {author} {\bibinfo {author} {\bibfnamefont {J.}~\bibnamefont
  {Zak}},\ }\bibfield  {title} {\enquote {\bibinfo {title} {Berry's phase for
  energy bands in solids},}\ }\href {\doibase 10.1103/PhysRevLett.62.2747}
  {\bibfield  {journal} {\bibinfo  {journal} {Phys. Rev. Lett.}\ }\textbf
  {\bibinfo {volume} {62}},\ \bibinfo {pages} {2747--2750} (\bibinfo {year}
  {1989})}\BibitemShut {NoStop}%
\bibitem [{\citenamefont {Makeenko}(2005)}]{S:Makeenko2005}%
  \BibitemOpen
  \bibfield  {author} {\bibinfo {author} {\bibfnamefont {Yuri}\ \bibnamefont
  {Makeenko}},\ }\href@noop {} {\emph {\bibinfo {title} {Methods of
  Contemporary Gauge Theory}}}\ (\bibinfo  {publisher} {Cambridge University
  Press},\ \bibinfo {year} {2005})\BibitemShut {NoStop}%
\bibitem [{\citenamefont {Duca}\ \emph {et~al.}(2015)\citenamefont {Duca},
  \citenamefont {Li}, \citenamefont {Reitter}, \citenamefont {Bloch},
  \citenamefont {Schleier-Smith},\ and\ \citenamefont
  {Schneider}}]{S:Duca2015}%
  \BibitemOpen
  \bibfield  {author} {\bibinfo {author} {\bibfnamefont {L.}~\bibnamefont
  {Duca}}, \bibinfo {author} {\bibfnamefont {T.}~\bibnamefont {Li}}, \bibinfo
  {author} {\bibfnamefont {M.}~\bibnamefont {Reitter}}, \bibinfo {author}
  {\bibfnamefont {I.}~\bibnamefont {Bloch}}, \bibinfo {author} {\bibfnamefont
  {M.}~\bibnamefont {Schleier-Smith}}, \ and\ \bibinfo {author} {\bibfnamefont
  {U.}~\bibnamefont {Schneider}},\ }\bibfield  {title} {\enquote {\bibinfo
  {title} {{An Aharonov-Bohm interferometer for determining Bloch band
  topology}},}\ }\href {\doibase 10.1126/science.1259052} {\bibfield  {journal}
  {\bibinfo  {journal} {Science}\ }\textbf {\bibinfo {volume} {347}},\ \bibinfo
  {pages} {288--292} (\bibinfo {year} {2015})}\BibitemShut {NoStop}%
\bibitem [{\citenamefont {Zenesini}\ \emph {et~al.}(2010)\citenamefont
  {Zenesini}, \citenamefont {Ciampini}, \citenamefont {Morsch},\ and\
  \citenamefont {Arimondo}}]{S:Zenesini2010}%
  \BibitemOpen
  \bibfield  {author} {\bibinfo {author} {\bibfnamefont {A.}~\bibnamefont
  {Zenesini}}, \bibinfo {author} {\bibfnamefont {D.}~\bibnamefont {Ciampini}},
  \bibinfo {author} {\bibfnamefont {O.}~\bibnamefont {Morsch}}, \ and\ \bibinfo
  {author} {\bibfnamefont {E.}~\bibnamefont {Arimondo}},\ }\bibfield  {title}
  {\enquote {\bibinfo {title} {{Observation of St\"uckelberg oscillations in
  accelerated optical lattices}},}\ }\href {\doibase
  10.1103/PhysRevA.82.065601} {\bibfield  {journal} {\bibinfo  {journal} {Phys.
  Rev. A}\ }\textbf {\bibinfo {volume} {82}},\ \bibinfo {pages} {065601}
  (\bibinfo {year} {2010})}\BibitemShut {NoStop}%
\bibitem [{\citenamefont {Kling}\ \emph {et~al.}(2010)\citenamefont {Kling},
  \citenamefont {Salger}, \citenamefont {Grossert},\ and\ \citenamefont
  {Weitz}}]{S:Kling2010}%
  \BibitemOpen
  \bibfield  {author} {\bibinfo {author} {\bibfnamefont {Sebastian}\
  \bibnamefont {Kling}}, \bibinfo {author} {\bibfnamefont {Tobias}\
  \bibnamefont {Salger}}, \bibinfo {author} {\bibfnamefont {Christopher}\
  \bibnamefont {Grossert}}, \ and\ \bibinfo {author} {\bibfnamefont {Martin}\
  \bibnamefont {Weitz}},\ }\bibfield  {title} {\enquote {\bibinfo {title}
  {{Atomic Bloch-Zener Oscillations and St\"{u}ckelberg Interferometry in
  Optical Lattices}},}\ }\href {\doibase 10.1103/PhysRevLett.105.215301}
  {\bibfield  {journal} {\bibinfo  {journal} {Phys. Rev. Lett.}\ }\textbf
  {\bibinfo {volume} {105}},\ \bibinfo {pages} {215301} (\bibinfo {year}
  {2010})}\BibitemShut {NoStop}%
\end{thebibliography}
